\RequirePackage{etoolbox}
\documentclass[ba]{imsart}
\pdfoutput=1

\pubyear{2019}
\volume{Authors' accepted version}
\doi{10.1214/19-BA1158}

\usepackage{amsthm}
\usepackage{amsmath}
\usepackage{natbib}
\usepackage[colorlinks,citecolor=blue,urlcolor=blue,filecolor=blue,backref=page]{hyperref}
\usepackage{graphicx}
\usepackage{amsfonts}
\usepackage{lscape}
\usepackage{multirow}
\usepackage{mathtools}
\usepackage{dsfont}

\DeclareMathOperator{\Y}{\boldsymbol{Y}}
\DeclareMathOperator{\y}{\mathbf{y}}
\DeclareMathOperator{\by}{\boldsymbol{y}}

\DeclareMathOperator{\bbeta}{\boldsymbol{\eta}}

\DeclareMathOperator{\bvare}{\boldsymbol{\varepsilon}}

\DeclareMathOperator{\f}{\mathbf{f}}
\DeclareMathOperator{\mf}{\mathrm{f}}
\DeclareMathOperator{\bbf}{\boldsymbol{f}}

\DeclareMathOperator{\btheta}{\boldsymbol{\theta}}
\DeclareMathOperator{\x}{\mathbf{x}}

\DeclareMathOperator{\s}{\mathbf{s}}
\DeclareMathOperator{\Rho}{\mathcal{P}}
\DeclareMathOperator{\cov}{\mathrm{Cov}}
\DeclareMathOperator{\Ex}{\mathbb{E}}
\DeclareMathOperator{\Vx}{\mathbb{V}}

\DeclareMathOperator{\0}{\boldsymbol{0}}

\DeclareMathOperator{\diag}{\mathrm{diag}}
\DeclareMathOperator{\bell}{\boldsymbol{\ell}}

\DeclareMathOperator*{\argmax}{arg\,max}
\DeclareMathOperator{\nW}{\mathrm{W}}

\startlocaldefs
\numberwithin{equation}{section}
\theoremstyle{plain}

\endlocaldefs

\begin{document}

\begin{frontmatter}
\title{Additive multivariate Gaussian processes for joint species distribution modeling with heterogeneous data \\ \normalsize Accepted for publication in Bayesian Analysis}
\runtitle{To appear in Bayesian Analysis, DOI: 10.1214/19-BA1158}
\thankstext{T1}{First available in Project Euclid: 3 June 2019: Available in Project Euclid: https://projecteuclid.org/euclid.ba/1559548823}

\begin{aug}
\author{\fnms{Jarno} \snm{Vanhatalo}\thanksref{addr1}\ead[label=e1]{jarno.vanhatalo@helsinki.fi}},
\author{\fnms{Marcelo} \snm{Hartmann}\thanksref{addr2}\ead[label=e2]{marcelo.hartmann@helsinki.fi}}
\and
\author{\fnms{Lari} \snm{Veneranta}\thanksref{addr3}}
\ead[label=e3]{lari.veneranta@luke.fi}
\runauthor{Vanhatalo, Hartmann and Veneranta}

\address[addr1]{Department of Mathematics and Statistics and Organismal and Evolutionary Biology Research Program, University of Helsinki, 
    \printead{e1} 
}

\address[addr2]{Department of Mathematics and Statistics and Department of Computer Science, University of Helsinki,
    \printead{e2}
}

\address[addr3]{Natural Resources Institute Finland, Finland,
    \printead{e3}}

\end{aug}

\begin{abstract}
Species distribution models (SDM) are a key tool in ecology, conservation and management of natural resources.
Two key components of the state-of-the-art SDMs are the description for species distribution response along environmental covariates and the spatial random effect that captures deviations from the distribution patterns explained by environmental covariates.
Joint species distribution models (JSDMs) additionally include interspecific correlations which have been shown to improve their descriptive and predictive performance compared to single species models.
However, current JSDMs are restricted to hierarchical generalized linear modeling framework.
Their limitation is that parametric models have trouble in explaining changes in abundance due, for example, highly non-linear physical tolerance limits which is particularly important when predicting species distribution in new areas or under scenarios of environmental change.
On the other hand, semi-parametric response functions have been shown to improve the predictive performance of SDMs in these tasks in single species models.

Here, we propose JSDMs where the responses to environmental covariates are modeled with additive multivariate Gaussian processes coded as linear models of coregionalization.
These allow inference for wide range of functional forms and interspecific correlations between the responses.
We propose also an efficient approach for inference with Laplace approximation and parameterization of the interspecific covariance matrices on the euclidean space.
We demonstrate the benefits of our model with two small scale examples and one real world case study.
We use cross-validation to compare the proposed model to analogous semi-parametric single species models and parametric single and joint species models in interpolation and extrapolation tasks.
The proposed model outperforms the alternative models in all cases.
We also show that the proposed model can be seen as an extension of the current state-of-the-art JSDMs to semi-parametric models.
\end{abstract}

\begin{keyword}[class=MSC]
\kwd[Primary ]{60G15}
\kwd{60K35}
\kwd[; secondary ]{62P12}
\end{keyword}

\begin{keyword}
\kwd{linear model of coregionalization}
\kwd{hierarchical model}
\kwd{heterogeneous data}
\kwd{spatial prediction}
\kwd{model comparison}
\kwd{Laplace approximation}
\kwd{covariance transformation}
\end{keyword}

\end{frontmatter}

\section{Introduction}

Species distribution models (SDMs) are key tools in the ecologists' toolbox. They have been widely used, among other applications, to study species habitat preferences \citep{Latimer+etal:2006}, to improve identification and management of conservation areas and natural resources \citep{meri:2016} and to evaluate species responses to environmental filtering under climate change scenarios \citep{clark+gelf:2014,Kottaetal:2019}. The main goals of statistical inference in these contexts are to use species observations and information on the associated environment to infer the relationship between these two attributes and to predict over regions of unsampled locations to build thematic species distribution maps \citep{gelf+sil:2006,elith}.

Species distribution modeling is, thus, directly related to inferring species' responses to environmental factors \citep{Latimer+etal:2006}. 
This is traditionally done using generalized linear or additive models as well as an array of machine learning approaches such as regression trees or maximum entropy modeling \citep{elith}.
These approaches model each species separately and cannot account for species interactions nor shared responses to the environment.
However, species interaction with other species is potentially as important factor as its response to environment. 
Moreover, in many practical situations, data from species can be patchy or scarce in which case sharing information between species can significantly improve models' predictive performance \citep{otso,clark+gelf:2014,Hui2013,thorson:2015,Hartmannetal:2017}. 
 For these reasons, joint species distribution models (JSDM) have gained increasing attention in recent years \citep{Warton2015}.

\citet{Dunstan2013} and \citet{Hui2013} introduced species archetype modeling where species' responses to the environment are clustered into few archetype models.
\cite{laura+reid:2014} proposed to use the multivariate probit regression model \citep{sidchib} to describe geographical co-occurrence patterns between frogs and eucalyptus trees in Australia and \cite{clark+gelf:2014} built a JSDM to infer richness and loss of species under climate change scenarios. 
\citet{thorson:2015} introduced a spatial latent factor model to predict spatial distribution of breeding birds and rock fish communities and recently, \cite{Ovaskainenetal:2017} introduced the hierarchical modeling of species communities (HMSC) framework which includes detailed description of interspecific correlations in covariate responses and spatial random effects.

Current JSDMs rely on the classical framework of hierarchical generalized linear models (GLMs). Even though this approach allows flexible modeling by describing the randomness of response variables with different probabilistic models \citep{nel+wed:1972}, it is still limited by its parametric assumptions. Hence, it may fail to accurately describe a species' response to environmental conditions \citep{vanh, Kottaetal:2019}.
Here, we propose a semi-parametric JSDM model represented with multivariate Gaussian processes (GPs). 
GPs are flexible semi-parametric regression models where the regression function is estimated without restrictive parametric assumptions about its form \citep{ohagan78,Rasmussen+Williams:2006}.  
Our model integrates the main strengths of semi-parametric single species models \citep{vanh,golding:2016} and generalized linear model based JSDMs. 

First, we model the species responses to environmental covariates with additive multivariate GPs. 
This allows us to capture wide range of nonlinear responses and share information about these responses between species. 
Second, due to the hierarchical model structure, the model can simultaneously accommodate several kinds of outcome variables (observations/measurements) by assigning different types of probabilistic models for them. 
This is important, since it allows us to exploit different types of measurements which commonly arise in real-case scenarios of multiple species surveys. Our presentation focuses in the mostly used probabilistic models in practice, the Bernoulli (Binomial) and Poisson with over-dispersion (Negative-Binomial).
We also propose an efficient computational approach build around Laplace method \citep{golding:2016,Vanhatalo+Pietilainen+Vehtari:2010}.
 Lastly, we present a structured cross-validation scheme to validate and compare models' performance in different kinds of prediction tasks.

This paper is organized as follows. In Section \ref{sec:caseStudy} we introduce a motivating case study. In section \ref{sec:additiveMVGP}, we introduce the additive multivariate GPs and in Section \ref{sec:posterior_inference} we discuss its predictive properties and introduce the computational methods for inference. In Section \ref{sec:experiments}, we illustrate the basic properties of the model through two simple examples and introduce the specific case study model. 
In Section \ref{sec:results} we present the case study results and 
we end by discussion and conclusions in Section \ref{sec:Discussion}.

\section{Motivating case study: coastal species distributions in the Gulf of Bothnia}\label{sec:caseStudy}

\begin{figure}[ht]
\setlength{\parindent}{-0.5cm}
\center
\includegraphics[scale = 0.3]{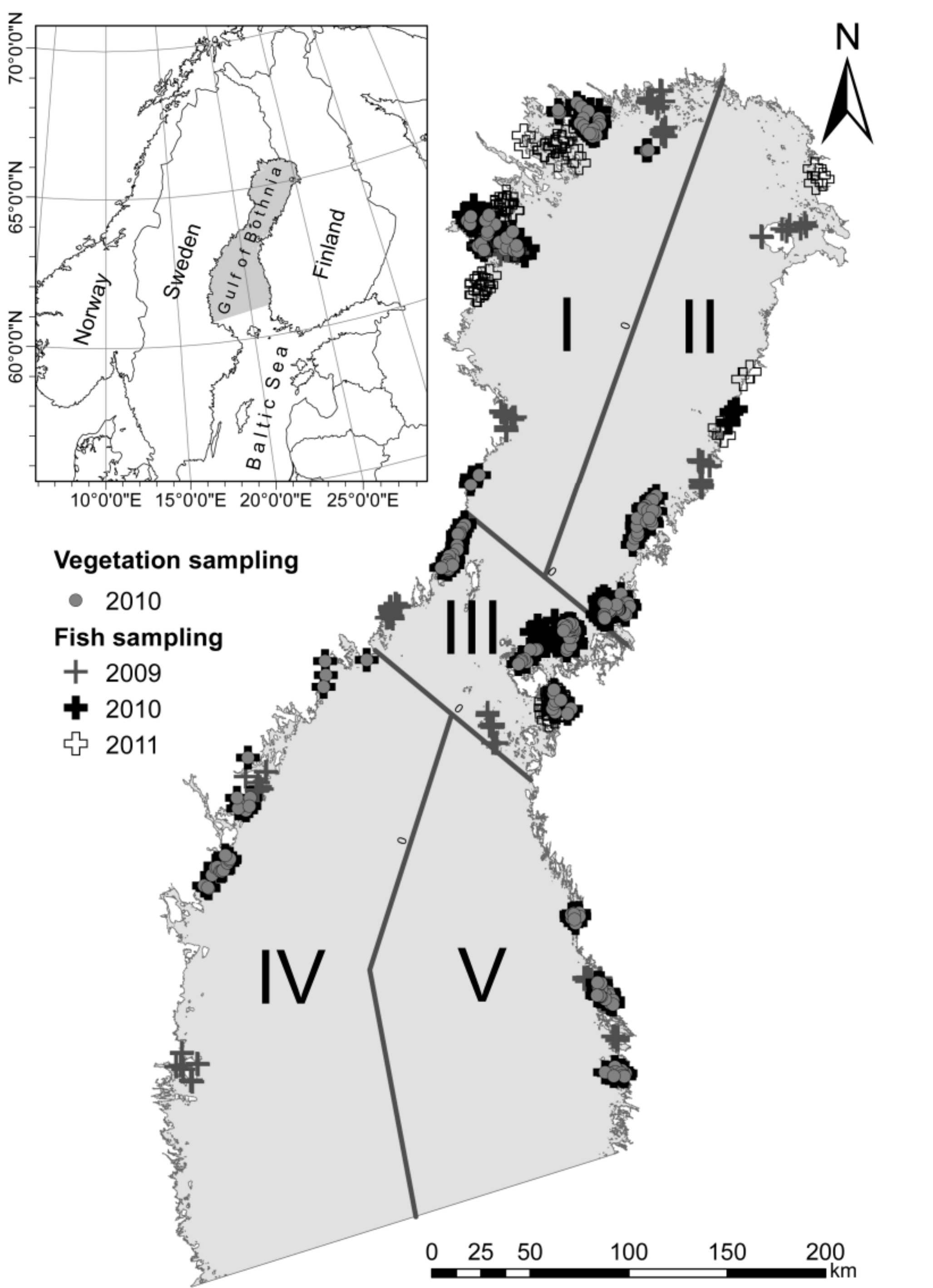}
\vspace{-0.3cm}
\caption{The study region, Gulf of Bothnia, and locations for species data. The region is divided into five subregions (I-V) to be used in cross-validation. The environmental conditions (described by the environmental covariates) are heterogeneous between these regions \citep[][]{lari:2013} which allows to test models' extrapolation performance.  
}
\label{fig:fig_3}
\end{figure}

As a motivating example, we study spring distribution of four fish species and three types of algae or macro-vegetation on the coastal region of the Gulf of Bothnia in the northern Baltic Sea.
The Gulf of Bothnia is a brackish water basin between Sweden and Finland covering an area of approximately $600 \times 120 \ \mathrm{km}$. 
Its coastal areas play a central role in the ecosystem and many Baltic fish stocks are dependent on the coastal regions for their reproduction \citep{lari:2013,meri:2016}.
  Coastal zones are also the most sensitive regions of the Baltic sea to both natural variation and anthropogenic pressures \citep{Reuscheaar8195}.  
  Hence, these areas are of central importance for conservation and there is need for detailed knowledge on the distribution of coastal species and predictions concerning their response to environmental changes.
  
\subsection{Case study species and data}
The studied fish species are the juvenile or adult three-spined stickleback (\emph{Gasterosteus aculeatus}) and nine-spined stickleback (\emph{Pungitius pungitius}) as well as larvae of  whitefish (\emph{Coregonus lavaretus}) and vendace (\emph{Coregonus albula}). 
Both whitefish and vendace are commercially important species and the sticklebacks are one key fish fauna in the Gulf of Bothnia ecosystem \citep{Bergstrometal:2015}. 
We treat the studied vegetation and algae in functional group level comprising of diatomous algae, filamentous and macro-vegetation. 
These are the dominating groups of benthic vegetation in shores where larvae of Coregonids (whitefish and vendace) occur at early developmental stages. 
In the scale of Gulf of Bothnia, their occurrence also reflects the salinity, nutrient balance and wind exposure of studied area \citep{lari:2013}. 

The case study species reflect the large scale environmental gradients and the changes in the Baltic Sea environment in last decades. 
Sticklebacks in the Baltic Sea use the shallow coastal zone for reproduction \citep{Bergstrometal:2015} 
and high abundances of sticklebacks in the Baltic Sea have been positively correlated with high occurrence of filamentous algae \citep{Eriksonetal:2009,Eriksonetal:2012}. 
Coregonids in coastal areas prefer more oligotrophic waters \citep{lari:2013}. 
Whitefish and vendace reproduce in stony reefs and islets of Baltic Sea in late autumn and the larvae hatch at ice break-up in spring. 
In sheltered areas the overwintering reeds (\emph{Phragmites australis}) dominate the macro-vegetation in spring. 
Diatomous algae consist of several epiphytic species that have an early spring bloom at ice break up and settle rapidly over bottom when light and water temperature increase \citep{Busse+Snoeijs:2002}. 
Filamentous algae in this study consist mostly of \emph{Pilayella sp}. 
It is a fast growing annual species that dominates the exposed shores at Baltic Sea in early spring  \citep{Ronnberg+Bonsdorff:2004}.

The whole study area was divided into sampling subareas (Figure~\ref{fig:fig_3}). 
Within each sub-area, data were collected at several sampling sites distributed so that they covered the range of most important environmental covariates in that sub-area.
Sampling was conducted in 2009-2011. At each site sampling was done approximately one week after the ice break.
In the scale of Gulf of Bothnia, the ice break-up happens in a period of approximately one month.
The species data used in this work comprise of 160 distinct sampling sites. 
In 70 sites (2010 data) all species were sampled and the rest of the sites comprise samples for the fishes only (Figure~\ref{fig:fig_3}). 
Fish samples comprise the number of caught fish together with information on the sampling effort at each site. 
The effort was measured as the volume of sampled water (m$^3$). A more detailed description of the data collection is provided by \citet{lari:2013}
For plant data, the bottom was photographed at 5-13 locations at depth of 30 cm within a distance of 2 m and parallel to shore line.
The occurrence of a species was recorded at 16 uniformly distributed points in all these photographs. 
The case study plants can grow over each others so that presence of one plant at a spatial location does not exclude the presence of other plants. 
The whitefish and vendace data were previously analyzed by \citet{lari:2013} and \citet{vanh} but the data on other species are unpublished.

\subsection{Environmental covariates}

\begin{table}[t]
\begin{tabular}{lll}
\hline
Variable            & Description  & Resolution {[}m{]} \\ \hline
Openness    & The average openness and exposure to winds & 300                \\
Distance to deep       & Distance to 20 m depth curve & 200                 \\
Sandy bottom index      & Area weighted distance to the sandy shores   & 90 \\
Ice breakup week    & The end date (weeks) of ice cover in 2009  & 1852     \\
Chlorophyll-a    & Chlorophyll-a concentration & 1852             \\
River     & Distance to the nearest river mouth & 150                       \\
Bottom class     & Bottom classification to 6 categories     & 200          \\
Winter salinity     & Length of ice winter & 10,000                          \\
\hline
\end{tabular}
\caption{\label{tab:environmental_Covars} Environmental covariates used in the case study and their original resolution. For this study, the resolution of all the environmental covariates was scaled to 300m.}
\end{table}

Gulf of Bothnia hosts rich variety of environmental conditions.
Coastal areas are affected by inflows from land as well as shallow and complex topography. 
In the scale of Gulf of Bothnia, there is a gradient in river influence, salinity, temperature and length of ice cover period from north to south. 
We used seven real-valued and one categorical abiotic environmental covariates that were available throughout the study area as raster maps.
Each species observation was combined with covariates from the raster cell where the sample location fell in. 
These are summarized in table \ref{tab:environmental_Covars} and described and motivated in detail by \citet{lari:2013}. 
Due its fresh water origin vendace is \emph{a priori} sensitive to even small changes in salinity levels experienced in the Gulf of Bothnia whereas the other species respond to salinity only in the Baltic Sea scale. 
Hence, we used all covariates only for vendace but excluded the winter salinity from other species.

\section{Hierarchical multivariate species distribution model}\label{sec:additiveMVGP}
We start our model building following the generic hierarchical structure similar to the one presented by, e.g., \cite{Wikle:2003}, \citet{cressie+wikle:2011} and \cite{Banerjee+Carling+Gelfand:2015} %
\begin{align}
\mathrm{[data | process, parameters]:} \hspace{0.2cm} & \pi_{\Y}(\by(\x,\s)|\f(\x,\s), \bbeta) \nonumber \\
\mathrm{[process | parameters]:} \hspace{0.2cm} & \pi_{\bbf}(\f(\x,\s)|\btheta) \nonumber \\
\mathrm{[parameters]:} \hspace{0.2cm}& \pi(\bbeta, \btheta)
\end{align}
where the first layer of hierarchy is the probabilistic model which defines the conditional distribution for multivariate data $\Y(\x,\s)$, at spatial location $\s$ with associated covariates $\x$, given the model parameters $\bbeta$ and the multivariate latent process $\bbf(\x,\s)$. 
The second layer defines the prior distribution for the latent process given the process parameters $\btheta$, 
and the third layer defines the prior distribution for all unknown parameters. 

Let $j \in \mathcal{J} = \lbrace 1, \cdots, J \rbrace$ be the species index set and $J$ the total number of species in the study. 
Denote by $\Y(\x, \s)^T =[Y_1 \cdots Y_J]$ the $J$-variate random vector with components $Y_j = Y_j(\x_j, \s)$ related to the $j^{th}$ species at spatial location $\s^T = [s_{1} \ s_{2}]$ (coordinates) under the influence of environmental covariates $\x_j^T = [x_{j, 1} \cdots x_{j, c_j}]$ where $c_j$ is the number of environmental covariates for the species $j$. 
We denote by $\x$ the full vector of covariates and $\mathcal{X}$ the covariate space. 
The species specific covariates $\x_j$ are sub-vectors of $\x$ such that $\x_j\in \mathcal{X}_j$ where $\mathcal{X}_j \subset \mathcal{X}$ and $\mathcal{X}_j$ is $c_j$-dimensional. 
Here, we assume that given a multivariate latent process $\bbf(\x,\s) = [f_1(\x_1, \s)  \cdots f_J(\x_J, \s)]^T:\mathcal{X}\times \mathbb{R}^2\rightarrow \mathbb{R}^J$, the probabilistic model for $\Y = \Y(\x, \s)$ factorizes as follows 
\begin{equation}
\pi_{\Y}(\by(\x,\s)|\f(\x,\s), \bbeta) = \prod\limits_{j=1}^J \pi_{Y_j}(y_j|\mf_j, \eta_j)
\end{equation}
where $\mf_j = \mf_j(\x_j,\s)$ is fixed but an unknown realization of the $j$'th process with covariates $\x_j$ at the spatial location $\s$. $\eta_j$ is the vector of parameters of the probabilistic model $\pi_{Y_j}$ for the species $j$ and $\bbeta = [\eta_1 \cdots \eta_J]^T$. 
%
In general any probabilistic model for data could be used and the observation models should be chosen according to the assumed sampling process. 
Here, we consider in detail the Binomial and the Negative-Binomial (over-dispersed Poisson) models used in our case study (Section~\ref{sec:caseStudy}). 
These models can be seen as observation models resulting from inhomogeneous point process model for species abundance \citep{gelf+sil:2006,Warton2010}.

In practice, each sampling site consists of a small area where the observations are made. In our case study, the area covered by a sampling site is so small compared to the whole study region that the latent function is practically constant in each sampling site. The model for the count of species presences at uniformly distributed points in a sampling site (plants in our case study) is then Binomial given by \citep{gelf+sil:2006}
\begin{equation}\label{bin}
\pi_{Y_B}(y|f(\x, \s), z_B(\s)) = {{z_B(\s)}\choose{y}} p(f(\x, \s))^{y} [1 - p(f(\x, \s))]^{z_B(\s) - y}I_{\lbrace 0, \hdots, z_B(\s) \rbrace}(y)
\end{equation}
where $z_B(\s)$ is the total number of observation points in the sampling site at location $\s$ and $p(f(\x, \s))$ is the success probability, which will be modeled through the logit here.
The natural model for the number of individuals in a closed area or volume in a sampling site (fish in our case study) is Poisson which we extend to over-dispersed Poisson using the Negative-Binomial given by \citep[see][Section 7.1 for details]{Liu+Vanhatalo:2019} 
\begin{equation}\label{negbin}
\pi_{Y_N}(y|f(\x, \s), z_N(\s), r) = \dfrac{\Gamma(r + y)}{y!\Gamma(r)}\left(\dfrac{r}{r + m(\x, \s)}\right)^r \left(\dfrac{m(\x, \s)}{r + m(\x, \s)} \right)^{y} I_{\mathbb{N}_0} (y)
\end{equation}
where $m(\x, \s)$ $=$ $z_N(\s) \exp[f(\x, \s)]=E (Y_N|f(\x, \s), z_N(\x), r)$ and $r$ is the over-dispersion parameter. 
The latent process $f(\cdot)$ corresponds to the logarithm of a species (relative) density and $z_{_N}(\s)$ is the sampled volume of water in the sampling site at location $\s$. 
For this parameterization $V(Y_N|f(\x, \s), z_N(\x), r)$ $=$ $m(\x, \s)$ $+$ $m(\x, \s)^2/r$. Increasing $r$ decreases the variance and when $r \rightarrow \infty$, the model approaches the Poisson distribution.

\subsection{Univariate additive latent Gaussian process}
First, we assume that marginally for any species $j$, the process model is given by
%
\begin{equation}\label{model:linear}
f_j(\x_j, \s) = \beta_{j, 0} + h_j(\x_j) + \varepsilon_j(\s)
\end{equation}
where $\beta_{j, 0}$ is the offset weight with distribution $\beta_{j, 0}|\sigma^2_{j, 0} \sim \mathcal{N}(0, \sigma^2_{j, 0})$ and $h_j : \mathcal{X}_j \rightarrow \mathbb{R}$ is a \emph{predictor function} of environmental covariates. 
The predictor function is assumed to be additive over the covariates, $h_j(\x_j)=\sum_{r=1}^{c_j}h_{j,r}(x_{j,r})$, and each additive term is given an independent zero mean GP prior, so that 
\begin{equation}
h_j(\x_j)|\btheta_j \sim \mathcal{GP}\left(0, \sum_{r=1}^{c_j}k_{h_{j,r}}(x_{j,r}, x'_{j,r}; \btheta_{j, r})\right)
\end{equation}
where $k_{h_{j,r}}(x_{j,r}, x'_{j,r}; \btheta_{j,r}) = \cov(h_{j,r}(x_{j, r}), h_{j,r}(x'_{j,r})|\btheta_{j, r})$ is a covariance function with parameter $\btheta_{j, r}$ and $\btheta_{j}$ $=$ $[\btheta_{j, 1}^T\cdots\btheta_{j, c_j}^T]^T$. For example, 
with $k_{h_{j,r}}(x_{j,r}, x'_{j,r}; \theta_{j,r})$ $=$ $x_{j,r}$ $x'_{j,r}\sigma^2_{j,r}$ and $\theta_{j,r}$ $=$ $\sigma^2_{j,r}$, the predictor function corresponds to linear model $h_{j,r}(x_{j,r})= x_{j,r}\beta_{j,r}$ where $\beta_{j,r}\sim N(0,\sigma^2_{j,r})$ \citep{Rasmussen+Williams:2006}. With other choices of covariance functions we can model non-linear responses in which case the model can be seen as an alternative to the traditional generalized additive models \citep[GAMs,][]{hastie:1986}. The general form of an additive GP prior would include also joint effects of covariates \citep{david:2011} but this is not considered here.

The term $\varepsilon_j(\s)$ is a spatial GP, 
\begin{equation}\label{randeff}
\varepsilon_j(\s)|\sigma^2_j, \bell_j \sim \mathcal{GP} \big(0, k_{\epsilon_j}(\s, \s'; \bell_j, \sigma^2_j) \big)
\end{equation}
where $k_{\epsilon_j}(\s, \s'; \bell_j, \sigma^2_j)$ is a spatial covariance function with variance $\sigma^2_j$ and range parameter $\bell_j$. 
When we consider independent models for each species, that is the traditional single species SDMs (SSDMs), the processes $h_{j,r}$ and $\epsilon_j$ are mutually independent among all species. The model outlined this far is similar to the GP based species distribution models used by, e.g., \citet{vanh}, \citet{meri:2016} and \citet{golding:2016}. Next, we introduce models which consider dependency.

\subsection{Additive multivariate Gaussian process priors}

In order to account for possible species interdependence, we first include spatial species interaction into the model through the linear model of coregionalization (LMC) \citep{mardia, Gelfand+Schmidth+Banerjee+Sirmans:2004}. Write $\bvare(\s)^T = [\varepsilon_1(\s) \cdots \varepsilon_J(\s)]$ and assume that $\bvare(\s)$ has LMC covariance structure with species specific correlation functions $\tilde{k}_{\epsilon_j}(\cdot, \cdot \ ; \bell_j)$. 
The covariance structure of the LMC with interspecific spatial dependence is then,
\begin{equation}\label{cov:lmc}
\cov\big(\varepsilon_j(\s), \varepsilon_{j'}(\s')| \Sigma_{\epsilon}, \bell \big) = \sum_{l = 1}^J u_{\epsilon,l}(j, j')\tilde{k}_{\epsilon_l}(\s, \s';\bell_l)
\end{equation}
with $\bell^T = [\bell_1^T \cdots \bell_J^T]$ and $u_{\epsilon,l}(j, j')$ the entry $(j, j')$ of $U_{\epsilon, l} = L_{\epsilon, l} L^T_{\epsilon, l}$ where $L_{\epsilon, l}$ is the $l^{th}$ column of the Cholesky decomposition $L_{\epsilon}$ of the coregionalization matrix $\Sigma_{\epsilon}$, that is matrix of interspecific correlations between spatial GP.
Hence, the vector-valued latent process $\bbf(\x, \s)^T = [f_1(\x_1, \s) \cdots f_J(\x_J, \s)]$ unconditional on $\beta_{1, 0}$,$\hdots$,$\beta_{J, 0}$ follows a multivariate GP which we denote as
\begin{equation}\label{gp:dpd_1}
\bbf(\x, \s)|\Lambda_1 \sim \mathcal{MGP}\bigg(\hspace{-0.1cm}\0 , \Sigma_0 + \sum_{r=1}^{c}k_{h_{r}}(\x_{r}, \x'_{r}; \btheta) + \sum_{j = 1}^J \tilde{k}_{\epsilon_j}(\s, \s';\bell_j)U_{\epsilon, j}\bigg)
\end{equation}
where $\Lambda_1 = (\Sigma_0, \Sigma_\epsilon, \btheta, \bell)$, $\Sigma_0 = \diag(\sigma^2_{1, 0}, \ldots, \sigma^2_{J, 0})$, $\btheta^T = [\btheta_1^T \cdots \btheta_J^T]$ and $k_{h_r}(\x_r, \x'_r; \btheta) = \diag\big(k_{h_{1,r}}(x_{1, r}, x'_{1, r}; \btheta_{1, r}), \cdots,k_{h_{J, r}}(x_{J, r}, x'_{J, r}; \btheta_{J, r})\big)$. If the predictor functions, $h_{j,r}$, were linear, the prior \eqref{gp:dpd_1} would correspond to traditional multivariate spatial model with independent linear effects and spatial LMC \citep{Gelfand+Schmidth+Banerjee+Sirmans:2004}.
%


We extend \eqref{gp:dpd_1} to an \emph{additive multivariate GP prior} where the dependence between the species specific additive predictor functions is modeled with LMC. 
We demonstrate this with a model where the set of covariates is equal for all species, that is, $\mathrm{dim}(\mathcal{X}_j)=c$ and $\x_j=\x$ for all $j$. Then the model is written as 
%
\begin{align}\label{gp:dpd_3}
\bbf(\x, \s)|\Lambda_2 & \sim \mathcal{MGP} \bigg(\hspace{-0.1cm} \0 , \Sigma_0 + \sum_{r = 1}^{c} \sum_{j = 1}^J  \tilde{k}_{h_{_{j, r}}}(x_{r}, x'_{r}; \btheta_{j, r}) U_{h_{_{r}}, j} + \sum_{j = 1}^J \tilde{k}_{\epsilon_j}(\s, \s';\bell_j)U_{\epsilon, j}\bigg) 
\end{align}
%
where $\Lambda_2 = (\Lambda_1, \Sigma_{h_1},\dots,\Sigma_{h_c})$ and $\Sigma_{h_r}$ is the interspecific covariance matrix between the species specific predictor functions of $r$'th covariate, 
$\tilde{k}_{h_{j, r}}(\cdot, \cdot; \btheta_{j, r})$ is a correlation function related to the predictor function $h_{j, r}$, $U_{{h_r}, j}= L_{h_{r}, j} L^T_{h_{r}, j}$ and $L_{{h_r}, j}$ is the $j^{th}$ column of the Cholesky decomposition $L_{h_r}$ of $\Sigma_{h_r}$. 
When the set of covariates is not the same for all species, covariate specific predictive functions are correlated only between species that share those same covariates. 

A JSDM with multivariate GP prior \eqref{gp:dpd_3} can be interpreted as extension of GP based SSDMs \citep{vanh,golding:2016} to joint species modeling similarly as done with the generalized linear model based JSDMs \citep{laura+reid:2014,Ovaskainenetal:2017}. 
The enhanced inferential ability of the multivariate additive GP compared to univariate GP models lies in its capability to infer similarity (dissimilarity) in species specific responses to covariates and in the spatial random effect. 
The similarity/dissimilarity of responses of two species $j$ and $j'$ along a covariate $r$ is indicated by a positive/negative value for correlation and hence, examining the LMC covariance matrices, $\Sigma_{h_r}$, can provide new insight to species to species associations. 


\subsection{Marginally uniform priors for correlation parameters}


Here, we define the prior for coregionalization covariance matrices but define the covariance functions and rest of the priors in Section \ref{sec:experiments}.
A standard choice for prior for correlation matrices is the inverse Wishart distribution. However, if there is no prior information on interspecific correlations, we follow \citet[][]{Hartmannetal:2017} and suggest to use marginally uniform priors for the LMC covariance matrices $\Sigma_{\epsilon}$ and $\Sigma_{h_{r}}$. These can be achieved by the distribution of \cite{barnard} and \cite{tokuda:2012}. Let $\Rho$ be a correlation matrix with elements $\rho_{j, j'}$. A prior distribution that is marginally noninformative for the correlations, that is, the marginal distribution for every $\rho_{j, j'}$ is uniform over $(-1, 1)$, is achieved with the distribution
\begin{equation}\label{eq:rho_prior}
\pi( \Rho |v) = \dfrac{\Gamma(\frac{v}{2})^J}{\Gamma_J(\frac{v}{2})} \vert \Rho \vert^{\frac{1}{2}(v-1)(J-1)-1} \prod_{j = 1}^J \vert \Rho_{jj} \vert^{-\frac{v}{2}} \mathds{1}_{(0, \infty)}(\det \Rho)
\end{equation}
when $v = J - 1$. Here, $\Gamma_J(\cdot)$ is the multivariate gamma function and matrix $\Rho_{jj}$ is obtained by removing the $j^{th}$ column and $j^{th}$ row of $\Rho$. When $v$ increases, the probability density \eqref{eq:rho_prior} becomes increasingly concentrated around the origin. 

\section{Posterior inference and predictive properties}\label{sec:posterior_inference}

Given a set of species observations at locations $\s_{i_j}, i_j=1,\hdots,n_j$, respectively for each species $j=1,\hdots,J$, the likelihood can be written as 
\begin{equation}\label{eq:lik}
\pi(\y|\f, \bbeta) = \prod\limits_{j=1}^J \displaystyle \prod\limits_{i_j = 1}^{n_j}\pi_j(y_{j, i_j}| \mf_{j, i_j}, \eta_j)
\end{equation}
where $y_{j, i_j}$ is the $i_j$'th observation of species $j$ at the  $i_j$'th spatial location  $\s^T_{i_j}$ associated with a set of covariates $\x_{i_j} \in \mathcal{X}_j$. 
The observed vector $\y = [\y_1^T \cdots \y_J^T]^T$ with $\y_j^T = [y_{j,1} \cdots y_{j,n_j}]$ collects all the species observations and $\f = [\f_1^T \hdots \f_J^T]^T$, where $\f^T_j = [\mf_{j,1}\cdots \mf_{j,n_j}]$, collects the corresponding latent variables. Hence, the likelihood factorizes over the latent variables and the inference can be done similarly as with univariate GP models. 
Markov chain Monte Carlo (MCMC) would provide exact answers in the limit of large sample size but deterministic approximations such as Laplace approximation or expectation propagation algorithm have also been shown to provide accurate approximate inference for univariate GP models with much lower computational time \citep{nick+2008,rue,Vanhatalo+Pietilainen+Vehtari:2010}. In order to study the properties of the model, we will examine the Laplace approximation for the posterior of latent variables conditional on the hyperparameters.

\subsection{Posterior predictive inference conditional on hyperparameters}

\subsubsection{Posterior of latent variables}\label{sec:posterior_of_latents}


The Laplace's method approximates the conditional posterior of the latent function values $\f_{\ast}=\f(\s_{\ast},\x_{\ast})$ at the spatial location $\s_{\ast}$ with covariates $\x_{\ast}$ as\footnote{In case of Gaussian observation model, this equals to the true posterior density of $\f|\y, \bbeta, \Lambda$.}
\begin{align} \label{eq:postlp}
\pi(\f_{\ast} | \y, \bbeta, \Lambda) \approx \mathcal{N}\big(\hspace{-0.05cm}\f_{\ast}| C_{\ast,\f}C^{-1} \hat{\f},C_{\ast} - C_{\ast,\f}(C^{-1} + \nW)^{-1}C_{\f,\ast}\big)
\end{align}
where $\hat{\f}$ is the (conditional) maximum a posterior (MAP) estimate of latent variables, 
\begin{align}\label{eq:fhat}
\hat{\f}=&\argmax_{\f \in \mathbb{R}^{\sum_j n_j}} \ \sum\limits_{j = 1}^{J}\sum\limits_{i_j=1}^{n_j} \log \pi_j(y_{j, i_j}| \mf_{j, i_j}, \eta_j) + \log \mathcal{N}(\f|\0, C)
\end{align}
and $\nW$ is the Hessian matrix of the negative log-likelihood function evaluated at $\hat{\f}$. Here, $C$ is the prior covariance matrix of $\f$, $C_{\ast,\f}$ is the (prior) covariance matrix between elements of $\f_{\ast}$ and $\f$. $C_{\ast}$ is the prior covariance of $\f_{\ast}$. In case of  multivariate additive GP \eqref{gp:dpd_3} the prior covariance matrix is given by
\begin{align}\label{eq:covdata_1}
C =& \left[ \begin{array}{ccc}
\sigma_0^2 \mathrm{J}_{n_1}  & \cdots & 0 \\ 
\vdots & \ddots & \vdots  \\ 
0 & \cdots & \sigma_0^2 \mathrm{J}_{n_J}
\end{array} \right] + 
\sum_{r = 1}^{c} \sum_{j = 1}^{J}  
\left[ \begin{array}{ccc}
u_{h_{_{r}}, j}(1,1) \{\tilde{K}_{h_{j, r}}\}_{1,1} & \cdots & u_{h_{_{r}}, j}(1,J) \{\tilde{K}_{h_{j,r}}\}_{1,J} \\ 
\vdots & \ddots & \vdots \\ 
u_{h_{_{r}}, j}(J,1) \{\tilde{K}_{h_{j,r}}\}_{J,1}  & \cdots &  u_{h_{_{r}}, j}(J,J) \{\tilde{K}_{h_{j,r}}\}_{J,J}
\end{array} \right]
\nonumber \\[0.3cm]
\phantom{=}& + 
\sum_{j = 1}^{J}  
\left[ 
\begin{array}{ccc}
u_{\epsilon,j}(1,1) \{\tilde{K}_{\epsilon_j}\}_{1,1} & \cdots & u_{\epsilon,j}(1,J) \{\tilde{K}_{\epsilon_j}\}_{1,J} 
\\ 
\vdots & \ddots & \vdots  \\ 
u_{\epsilon,j}(J,1) \{\tilde{K}_{\epsilon_j}\}_{J,1}  & \cdots &  u_{\epsilon,j}(J,J) \{\tilde{K}_{\epsilon_j}\}_{J,J} 
\end{array} \right]
\end{align}
where $\mathrm{J}_n$ is an $n\times n$ matrix full of ones $u_{\epsilon,l}(j,j')$ and $u_{h_r,l}(j,j')$ are the entry $(j,j')$ of $U_{\epsilon,l}$ and $U_{h_r,l}$ (see \eqref{cov:lmc}), and $\{\tilde{K}_{\epsilon_l}\}_{j,j'}$ and $\{\tilde{K}_{h_{l,r}}\}_{j,j'}$ are correlation matrices with elements $[\{\tilde{K}_{\epsilon_l}\}_{j,j'}]_{i_j,i_{j'}} = \tilde{k}_{\epsilon,l}(\s_{i_j}, \s_{i_{j'}} \ ; \bell_l) $ and $[\{\tilde{K}_{h_{l,r}}\}_{j,j'}]_ {i_j,i_{j'}} = \tilde{k}_{h_{l,r}}(x_{i_j,r},x_{i_{j'},r}|\btheta_{l, r})$.
The other correlation matrices are constructed similarly. 
If data on all species are available at all sampling locations, the covariance matrix reduces to Kronecker product similarly as in LMC models by \citet{mardia} and \citet{Gelfand+Schmidth+Banerjee+Sirmans:2004}, so that
$C=\Sigma_0 \otimes J_n + \sum_{r = 1}^{c} \sum_{j = 1}^{J}  U_{h_{_{j, r}}, j} \otimes \tilde{K}_{h_{j, r}} + \sum_{j = 1}^{J} U_{\epsilon, j} \otimes \tilde{K}_{\epsilon_j} 
$
%
%
where $n_j=n$ for all $j=1,\hdots,J$. 
Since, in general, the second and third term of $C$ are full matrices, it can be seen from \eqref{eq:postlp} and \eqref{eq:fhat} that the posterior of $\f$ and the posterior predictive distribution of $\f_{\ast}$ are affected by observations of all species at any spatial locations. 

\subsubsection{The marginal posterior of additive terms}\label{sec:marginal_additive_affects}
Typically, we are also interested in species' responses along covariates encoded by the additive predictor functions. Their marginal posteriors, conditional on hyperparameters, are analogous to \eqref{eq:postlp}. The matrices $C_{\ast}$ and $C_{\ast,\f}$ are only constructed using the correlation functions corresponding to the latent function of interest. For example, to study the response of species $j$ along covariate $x_r$  we evaluate the posterior predictive distribution for $h_{j,r}(x_{r})$ using \eqref{eq:postlp} so that we replace $C_{\ast,\f}$
\begin{equation}\label{eq:pred_Chf}
C_{h_{j,r},\f} = \sum_{l=1}^J\left[u_{h_{r},l}(j,1) \{\tilde{K}_{h_{l, r}}\}_{j,1}, \cdots, u_{h_{_{r}}, l}(j,J) \{\tilde{K}_{h_{l,r}}\}_{j,J}\right],
\end{equation}
and similarly for $C_{\ast}$. However, in case of the traditional LMC \eqref{gp:dpd_1} only the spatial random effects are correlated between species whereas the predictor functions are not. 
Hence, $\Sigma_{h_r}$ is diagonal for all $r$ so that $u_{h_r,l}(j,j')=\sigma_l^2$ if $l=j=j'$ and zero otherwise. 
The second terms of \eqref{eq:covdata_1} is then block diagonal and only the $j$'th block column in $\eqref{eq:pred_Chf}$ is non-zero for which reason there is no information exchange between species specific predictor functions. 
In case of univariate GP prior all the terms in \eqref{eq:covdata_1} 
are block diagonal and \eqref{eq:postlp} reduces to $J$ independent Gaussian distributions with no information exchange among species at all. 
Hence, the predictive performance of the univariate GP and the two multivariate GP models are very different. This is illustrated in section~\ref{sec:experiments}.

\subsubsection{The (marginal) posterior expectation and variance for new observations}\label{sec:marginal_uniform_corr_prior}

When predicting species occurrence probability or abundance, we need to marginalize over the posterior of the latent variables. 
The posterior expectation and variance for a new outcome for species $j$ at a location $(\x_*, \s_*)$ conditional on hyperparameters are
{
\begin{align}\label{eq:predE}
\Ex[Y_j(\x_*, \s_*)|\y, \bbeta, \Lambda] =& \Ex [ \Ex (Y_j(\x_*, \s_*)|f_j(\x_*, \s_*), \y, \bbeta, \Lambda)] \\ 
\Vx[Y_j(\x_*, \s_*)|\y, \bbeta, \Lambda] =& \Ex[\Vx(Y_j(\x_*, \s_*)|f_j(\x_*, \s_*), \y, \bbeta, \Lambda)] \ + \nonumber \\ 
& \hspace{3cm} \Vx[\Ex(Y_j(\x_*, \s_*)|f_j(\x_*, \s_*), \y, \bbeta, \Lambda)] \nonumber
\end{align}
When the probabilistic model for species $j$ is assumed to be the Negative-Binomial or Binomial model with logistic link function \eqref{bin} we can find either an exact result or an analytical approximation for these expectations and variances. These are given in the \ref{suppA}.
These solutions speed up the predictive calculations considerably compared to numerically integrating $f$ over $\mathbb{R}$.

\subsubsection{Conditional scenario predictions}\label{sec:conditional_pred}

In some applications, one might be interested in scenario predictions conditional on changes in species composition. For example, one might be interested in how removing from or introducing specific species into an area would affect other species. 
In our case study setting, we could be interested in, for example, effects of management actions that would clean filamentous algae from the shoreline.
Such scenario predictions would be naturally tackled with predictive causal inference \citep{Lindley2002,Pearl:2009}.
In predictive causal inference the parameters of the model are inferred with the available data so far and predictions made considering alternative scenarios. 

%

To illustrate this lets first introduce a short notation $\y_{\mathcal{J}_1, *}$ $=$ $\y_{\mathcal{J}_1}(\x_*, \s_*)$ and $\bbf_{\mathcal{J}_1, *}$ $=$ $\bbf_{\mathcal{J}_1}(\x_*, \s_*)$ where $\mathcal{J}_1$ denotes the set of species to be predicted and $\mathcal{J}_2$ the \emph{scenario species} assumed to be ``managed''. For brevity, we omit the conditioning on hyperparameters and data. 
The conditional distribution of $\Y_{\mathcal{J}_1}(\x_*, \s_*)|\y_{\mathcal{J}_2}$ is 
\begin{align} \label{eq:condpred2}
\pi(\y_{\mathcal{J}_1, *}|\y_{\mathcal{J}_2}) 
&= \displaystyle\int
\pi(\y_{\mathcal{J}_1, *}| \y_{\mathcal{J}_2}, \f_{\mathcal{J}_1, *})\pi(\y_{\mathcal{J}_2}, \f_{\mathcal{J}_1, *}) \mathrm{d}\hspace{-0.04cm}\f_{\mathcal{J}_1, *}/ \pi(\y_{\mathcal{J}_2})
\end{align}
where $Y_{\mathcal{J}_1}(\x_*, \s_*)| \y_{\mathcal{J}_2}, \f_{\mathcal{J}_1, *}$ only depends on $\f_{\mathcal{J}_1, *}$. The Laplace approximation for this conditional predictive distribution is shown in \ref{suppA}.

\subsection{Parameter inference}

We used Laplace approximation \citep{Vanhatalo+Pietilainen+Vehtari:2010} to approximate the conditional posterior of latent variables $\pi(\f | \y, \bbeta, \Lambda)$ and the marginal likelihood of the hyperparameters $\pi(\y|\bbeta, \Lambda)=\int \pi(\y|\f,\bbeta)\pi(\f|\Lambda)d\f$.
We searched for the (approximate) maximum a posterior (MAP) estimate for hyperparameters given by
\begin{equation}
(\hat{\bbeta},\hat{\Lambda}) = \underset{\bbeta,\Lambda}{\argmax} \log q(\y|\bbeta, \Lambda) + \log \pi(\bbeta, \Lambda).
\end{equation}
where $\log q(\y|\bbeta, \Lambda)$ is the Laplace approximation for the log marginal likelihood for parameters. 
This approach produces also good approximation for the posterior predictive distribution for latent variables \citep{Vanhatalo+Pietilainen+Vehtari:2010,Vehtari2016}, which are the main interest in this work.
Hence, we fixed hyperparameters to their MAP estimate.

In order to avoid constrained optimization, all the parameters were transferred to unconstrained space. 
We used log transformation for covariance function parameters, and for the interspecific correlation matrices we used the transformation presented by \cite{kuro:2003} and \cite{lkj:2009}, which is a bijective mapping between the space of correlation matrices and $\mathbb{R}^{J\choose2}$. 
This is summarized in \ref{suppA}. 
We used scaled conjugate gradient optimization for locating the MAP and checked carefully that a (local) mode had really been found by verifying that gradients along all hyperparameters were zero. 
The required gradients of $\log q(\y|\bbeta,\Lambda)$ were solved analytically as described by \citet{Rasmussen+Williams:2006} and \citet{Vanhatalo+Pietilainen+Vehtari:2010}. 
The \ref{suppA} summarizes the derivatives w.r.t. to the covariance parameters in $\Sigma_{\epsilon}$, $\Sigma_{h_r}$, $r$ $=$ $1, \hdots, c$.


\section{Experiments}\label{sec:experiments}

Here, we first illustrate the properties of the hierarchical multivariate GP models with two simple examples. 
These examples highlight particular properties of the proposed model. 
After this we introduce the model and analysis for our  case study (Section~\ref{sec:caseStudy}). 
We implemented all the models using the GPstuff package \citep{Vanhatalo2013a}.

\subsection{Demonstration with simulated spatial data}\label{sec:spatial_demo}

Figure~\ref{fig:fig_2} presents simulated data and posterior predictions for spatial modelling of two species. 
The model follows spatial LMC; that is model \eqref{gp:dpd_1}, where the covariate terms are dropped out. The spatial correlation function used in this demo is the Gaussian $k(\s,\s') = e^{-||\s-\s'||^2/2 l^2}$, where $||\s-\s'||$ is the Euclidean distance and $l$ is the length-scale.
 The first row of subplots shows the posterior predictive probability of observing species 1 ($E[Y_1]$, $Y_1(\s) \sim \mathrm{Binomial}$) and the second row shows the expected number of species 2 ($E[Y_2]$, $Y_2(\s) \sim$ Negative-Binomial). 
Plots  (a) and (e) show the predictions when the species observations are from the same region but not from exactly the same locations. 
In this case, the prediction of multivariate GP is similar to predictions by univariate GPs.
Plots (b) and (f) show the predictions when data are available on both species from the lower left corner and additionally data on species 1 is available from upper right corner.
There is positive correlation between species, which has been inferred from the data in the lower left region, so the prediction for species 2 in upper right corner is informed by data on species 1 in that region. 
The last two columns illustrate the conditional scenario predictions \eqref{eq:condpred2}.
In the third column, the training data is the same as in the first column and the expected values in plots (c) and (g) show joint scenario prediction in new region of same size and shape. 
In this scenario, species in the new region were observed so that species 1 is known to be in the top right, and species 2 in the bottom left.
The fourth column shows the conditional scenario prediction for both of the species separately so that the grey marks show the locations where the other species would be introduced in these scenarios.

\begin{figure}[t]
\setlength{\parindent}{-0.45cm}
\includegraphics[height = 6cm, width = 14.2cm]{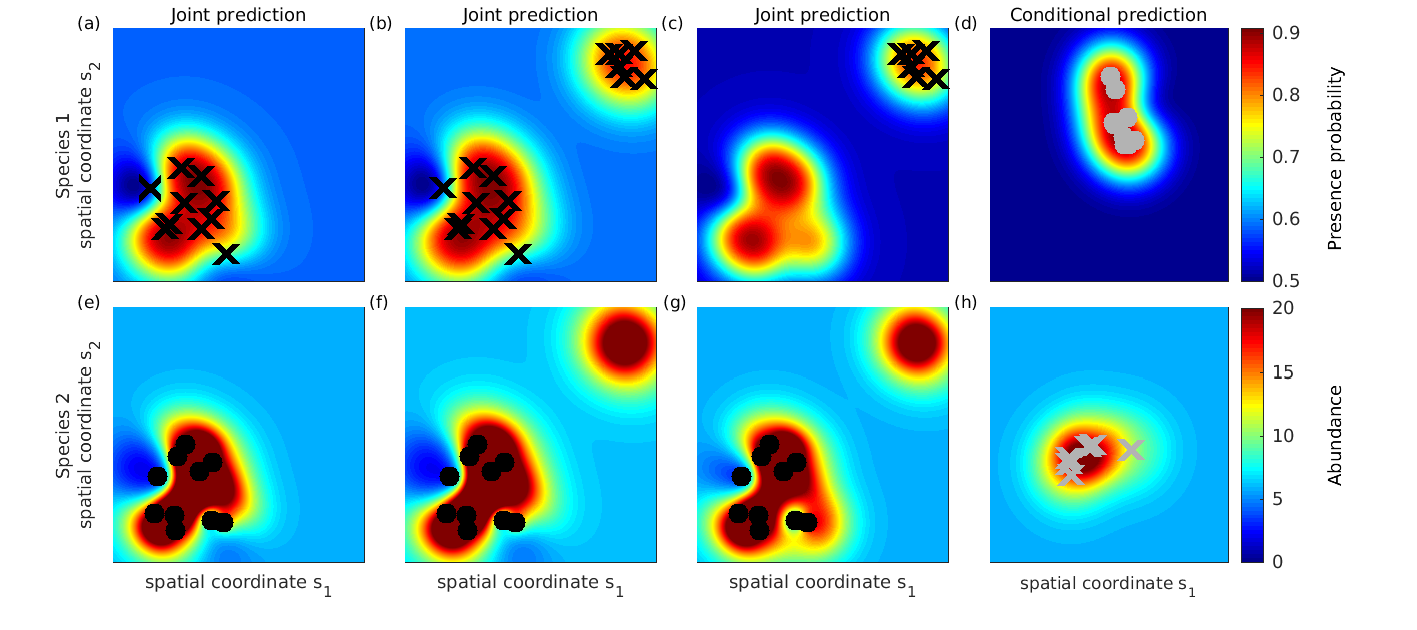}
\vspace{-1cm}
\caption{Illustration of spatial multivariate GP prior for JSDM with two species and spatial only component. Crosses and dots are simulated data locations of species 1 (Binomial data) and species 2 (Negative-Binomial data). See Section~\ref{sec:spatial_demo} for discussion.} 
\label{fig:fig_2}
\vspace{-0.4cm}
\end{figure}

\subsection{Demonstration with time series of species abundances}\label{sec:temporal_demo}

Next, we consider the classical predator-prey system containing annual population counts of hare and lynx in the northern Canada from 1845 to 1935 \citep{odum:1953}. 
The data is not spatially explicit since the observations are total counts over the region so we model it as time series. 
This illustrates the behavior of individual additive covariate effect in the multivariate GP model~\eqref{gp:dpd_3} with time being the covariate.
Let's denote by $Y_{1, t}$ and $Y_{2, t}$ the population sizes of hare and lynx at time $t$, and assume that $Y_{1, t}|f_1(t) \sim \mathrm{Poisson}(\exp(f_1(t))$ and $Y_{2, t}|f_2(t) \sim \mathrm{Poisson}(\exp(f_2(t))$. 
We compared two alternative priors for the latent functions, one with independent GP priors and another with joint bivariate GP prior with the LMC structure.
In order to compare models' predictive performances when some species have not been observed, we removed parts of the original data from the training set, the period of 1870-1900 for hare and 1850-1870 for lynx.  
Figure~\ref{fig:timeseries} displays the result of the data analysis. 
The model with bivariate GP prior clearly outperforms the independent model since it predicts better the test data points in periods where training data was removed. Moreover, its predictions have also smaller uncertainty than the predictions of independent GPs in those periods of time. 

\begin{figure}[t]
\setlength{\parindent}{-0.2cm}
\includegraphics[height = 7cm, width = 14cm]{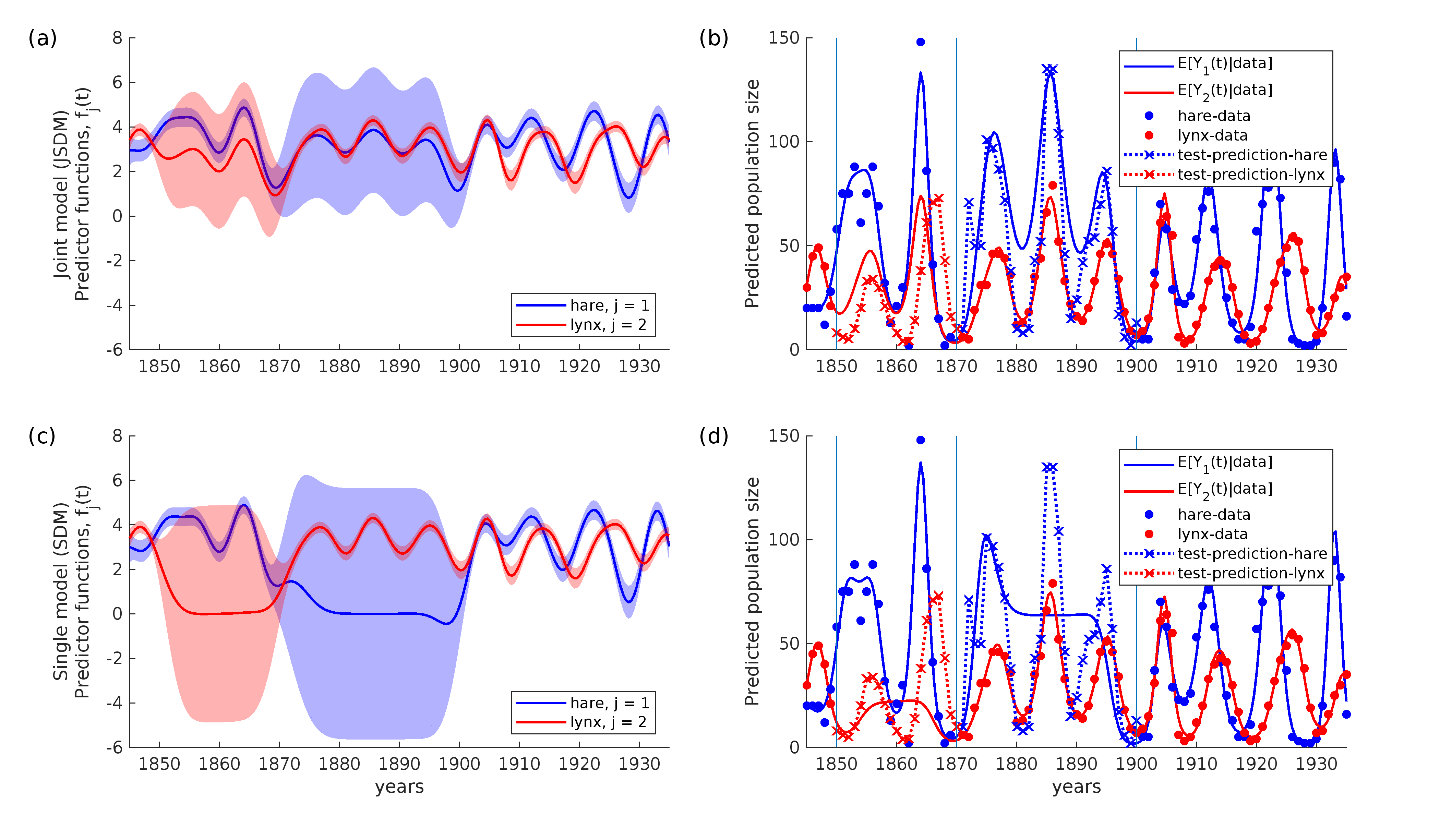}
\vspace{-1.1cm}
\caption{The results from the analysis of hare and lynx interaction (predator-prey system) analyzed with single and joint SDM. Plots a) and c) show the posterior mean and 95\% credible interval of the latent functions and plots b) and d) show the training data, test data and posterior expectation of observations.  See Section~\ref{sec:temporal_demo} for explanation.} 
\label{fig:timeseries}
\vspace{-.1cm}
\end{figure}

\subsection{The case study on coastal species distribution}

\subsubsection{Case study models}\label{sec:case_study_models}

The plant data are modeled with the Binomial observation model \eqref{bin} and the fish data are modeled with the Negative-Binomial observation model \eqref{negbin}. 
In order to test the effect of different model components and the effect of semi-parametric response functions versus standard quadratic response functions we compare the following models: 
\begin{itemize}\setlength\itemsep{-0.35em}
\item[1)] additive univariate GP predictor functions and univ. spatial random effects \eqref{model:linear}, 
\item[2)] additive univariate GP predictor functions and LMC spatial random effect \eqref{gp:dpd_1}, 
\item[3)] additive multivariate GP predictor function and LMC spatial random effect \eqref{gp:dpd_3}, 
\item[4)] additive univariate GP predictor functions only (model \eqref{model:linear} without $\epsilon_j(\s)$), 
\item[5)] univariate spatial random effects only (model \eqref{model:linear} without $h_j(\x_j)$), 
\item[6)] additive univariate quadratic predictors and univariate spatial random effects (model~\eqref{model:linear} with $h_j(\x_j)=\x_j\beta_j$ and $\x_j$ includes the covariates and their squares), 
\item[7)] additive univariate quadratic predictors and LMC spatial random effect (model~\eqref{gp:dpd_1} with $h_j(\x_j)=\x_j\beta_j$ and $\x_j$ includes the covariates and their squares), 
\item[8)] additive multivariate quadratic predictors and LMC spatial random effects (model~\eqref{gp:dpd_3} with $h_j(\x_j)=\x_j\beta_j$ and $\x_j$ includes the covariates and their squares).
\item[9)] additive multivariate quadratic predictors only (model~\eqref{gp:dpd_3} without $\epsilon(\s)$ with $h_j(\x_j)=\x_j\beta_j$ and $\x_j$ includes the covariates and their squares).
\end{itemize}
In each model, the spatial random effects were given the M\'{a}tern covariance function with $\nu=3/2$ degrees of freedom 
$k_{\epsilon_j}(\s, \s';\sigma^2_j, \bell_j) = \sigma^2_j \big(1 + \sqrt{3} \ r_j(\s, \s') \big) e^{-\sqrt{3} r_j(\s, \s')} $
where $\sigma_j^2$ is the variance parameter, $\bell_j^T = [\ell_{j, 1} \ \ell_{j, 2}]$ is the vector of length-scales and $r_j(\s, \s') = [(\s - \s')^T (\diag(\bell_j)^{2})^{-1}(\s - \s')]^{1/2}$.
The continuous covariate effects in the additive GP models (models 1-4) were given the Gaussian correlation function $\tilde{k}_{h_j,r}(x_{r},x_r') = e^{-||x_r-x_r'||^2/2 l_{j,r}^2}$. 
The additive linear models (models 6-8) were coded as additive GPs with 
 $\tilde{k}_{h_{j,r}}(x_{r}, x'_{r})=x_{r}x'_{r}\sigma^2_{j,r}$. 
 Even though this is not a proper correlation function, when used in LMC it implies a generalized linear model with interspecific dependencies between weights, $\beta_j$, that are the key components in state-of-the-art parametric JSDMs. 
 See Discussion for details.
For the categorical covariate (Bottom class, table \ref{tab:environmental_Covars}) we used a delta function so that $\tilde{k}_{h_{j,r}}(x_r, x_r'; \btheta) = \sigma_{\delta_j}^2 \delta_{x_r}(x_r')$, where $\delta_{x_r}(x_r')=1$ if $x_r=x_r'$ and zero otherwise. 
This corresponds to having an own intercept for each category.
We used the marginally uniform priors (section \ref{sec:marginal_uniform_corr_prior}) for the between species correlations and weakly informative priors for the rest of the parameters. The variance parameters were given $\mathrm{Half}\mbox{-}\mathrm{Student}\mbox{-}\it{t}$($\mu = 0$, $\sigma^2 = 4$, $\nu = 4$) priors and the length-scale parameters were given  $\mathrm{Half}\mbox{-}\mathrm{Inverse\mbox{-}Student}\mbox{-}\it{t}$($\mu = 0$, $\sigma^2 = 1$, $\nu = 4$). Hence, \emph{a priori} more weight is given for smooth functions with small variability.

\subsubsection{Model validation and comparison}

We aim to provide models that give reliable posterior predictions. Hence, it is natural to compare models with the goodness of their posterior predictive distributions. This can be done efficiently with cross-validation (CV) using the log predictive density diagnostics \citep{Vehtari+Ojanen:2012}. 
Let $\mathcal{D}$ denote the full data-set. Fix $K$ disjoint sub-sets of $\mathcal{D}$, say $\mathcal{D}_1, \ldots, \mathcal{D}_K $, such that their union is $\mathcal{D}$. 
The K-fold CV log point-wise marginal predictive density is then
 $L_K(\mathcal{D}) = \frac{1}{n}\sum_{i = 1}^{n} \log \pi(y_i|\mathcal{D}_{\sim k(i)})$
where $\mathcal{D}_{\sim k(i)} = \lbrace \cup_{r = 1}^K \mathcal{D}_r : r \neq k(i) \rbrace$ and $k(i)$ is such that $y_i\in \mathcal{D}_{k(i)}$. 
We compare the models with the leave-one-out CV (LOO-CV, $K = n$) and structured 5-fold CV. 
We used the MAP estimate for the hyperparameters. In 5-fold CV we found the MAP for each training set separately. 
The LOO-CV was conducted at the MAP found with full data and we used Laplace approximation to approximate the LOO-CV log predictive densities \citep{Vehtari2016}. 
Laplace approximation for LOO-CV is shown to work well in GP models and, since our data is rather large, leaving only one data point out of the training set has only negligible effect on the posterior of the hyperparameters \citep{Vehtari2016}.

The rationale for calculating both LOO-CV and 5-fold CV comparison is the following.
Since multiple species were sampled at every sampling site and each of the sampling sites has other sites spatially nearby it, the LOO-CV log predictive densities are affected significantly by the spatial random effects. The LOO-CV, thus, measures models' interpolation performance which can be good even if the models were not able to represent well responses along covariates (predictor functions) \citep{vanh,lari:2013}. 
For this reason, we structured the 5-fold CV by dividing the data into five spatially distinct groups corresponding to regions I-V in Figure~\ref{fig:fig_3}. 
The sampling sites in different groups are spatially so far from each others that the spatial random effects do not affect the posterior predictive distributions for test groups. 
Hence, the 5-fold CV tests mostly the extrapolation performance of a model, which is governed by the goodness of the predictor functions, whereas the LOO-CV tests the interpolation performance, which is governed also by the spatial random effects.

\begin{table}[t]
\begin{tabular}{l|cc}
\hline
GP models     & LOO Cross Validation  & 5-fold Cross validation \\ \hline
\multirow{1}{5.5cm}{1) Univariate GP $h(\x)$, univ. $\epsilon(\s)$ } &  -2.230 (0.082) $\phantom{-}$	& -2.465 (0.094) $\phantom{-}$  \\
\multirow{1}{5.5cm}{2) Univariate GP $h(\x)$, multiv. $\epsilon(\s)$ } &  -2.081 (0.080) $\phantom{-}$ & -2.480 (0.100) $\phantom{-}$ \\ 
\multirow{1}{5.5cm}{3) Multiv. GP $h(\x)$, multiv. $\epsilon(\s)$}   &  -\textbf{1.669 (0.080)} $\phantom{-}$ & \textbf{-2.316 (0.086)} $\phantom{-}$ \\
\multirow{1}{5.65cm}{4) Multiv. GP $h(\x)$ only } &   -2.228(0.089) $\phantom{-}$ &  -2.590(0.012) $\phantom{-}$ \\
\multirow{1}{5.65cm}{5) Multiv. $\epsilon(\s)$ only }  &  -2.351(0.087) $\phantom{-}$ & -2.500(0.086) $\phantom{-}$ \\
 \hline
\multirow{1}{6.1cm}{6) Univ. quadr. $h(\x)$, univ. $\epsilon(\s)$ } &  -2.262 (0.083) $\phantom{-}$	& -2.517 (0.095) $\phantom{-}$  \\
\multirow{1}{6.1cm}{7) Univ. quadr. $h(\x)$, multiv. $\epsilon(\s)$ } &  -2.117 (0.084) $\phantom{-}$ & -2.484 (0.109) $\phantom{-}$ \\ 
\multirow{1}{6.1cm}{8) Multiv. quadr. $h(\x)$, multiv. $\epsilon(\s)$ }   &  -2.105 (0.084) $\phantom{-}$ & -2.416 (0.099) $\phantom{-}$ \\
\multirow{1}{5.65cm}{9) Multiv. quadr. $h(\x)$ only}  &   -10.246(1.297) $\phantom{-}$	&  -9.305(1.008) $\phantom{-}$  \\
\end{tabular} 
\caption{\label{tab:loocv} Model comparison with leave-one-out (LOO) and 5-fold cross validation using the mean point wise log marginal predictive density statistics (and its standard error) over all species. see Section~\ref{sec:case_study_models} for model descriptions. }
\end{table}

\section{Results}\label{sec:results}

\subsection{Predictive performance of models}

Table~\ref{tab:loocv} summarizes the CV comparisons over all species and tables~1-2 in \ref{suppA} show the models' predictive performance for each species separately. 
Tables~3-4 in \ref{suppA} show the pairwise comparison of the CV point wise log predictive densities between the best GP/parametric model (models~3 and 8) and the other GP/parametric models with both the covariate effects and spatial random effect (models~1-2 and 6-7). 
The results show that model~3, which includes multivariate GP predictors and multivariate spatial random effects \eqref{gp:dpd_3}, is the best in both LOO-CV and 5-fold CV with a difference of $10^{-2}$ or more to the other models.
Since the mean log predictive density is an average of $n=850$ point-wise predictions the difference of $10^{-2}$ corresponds to a difference of 8.5 in log (point-wise) joint predictive densities. Analogously to Bayes factors, which compare log prior predictive distribution, this can be considered a significant difference between two models \citep{Kass+Raftery:1995}.

Hence, the multivariate GP (model~3) has significantly better overall predictive performance over all species than the other models. 
According to posterior predictive checks \citep{Gelman+Carlin+Stern+Rubin:2013} there are no serious discrepancies between its predictions and observed data.
In extrapolation (5-fold CV) model~3 performs best also for all species separately. 
However, in interpolation (LOO-CV) it is not the best for all species separately (tables~1-2 in \ref{suppA}).
Moreover, the semi-parametric GP models (models 1-4) work better than the corresponding parametric models  (quadratic environmental response, models 6-9) in both interpolation and extrapolation in general. 
Dropping either spatial random effect or covariate effect out from the model decreases its performance clearly.
All models work better in interpolation than in extrapolation and compared to univariate models including interspecific correlations either in spatial random effects or also in environmental predictors improves models' performance in both of these tasks. 
Moreover, including interspecific correlations between environmental responses, $h_{j,r}(\x_r)$, improves the extrapolation performance relatively more than including interspecific correlations only for spatial random effects. Hence, multivariate spatial random effect improved more interpolation whereas multivariate predictors has larger effect for extrapolation as would intuitively be expected.

\begin{figure}[!t]
\setlength{\parindent}{-0.45cm}
\includegraphics[width=13cm, height=5.3cm]{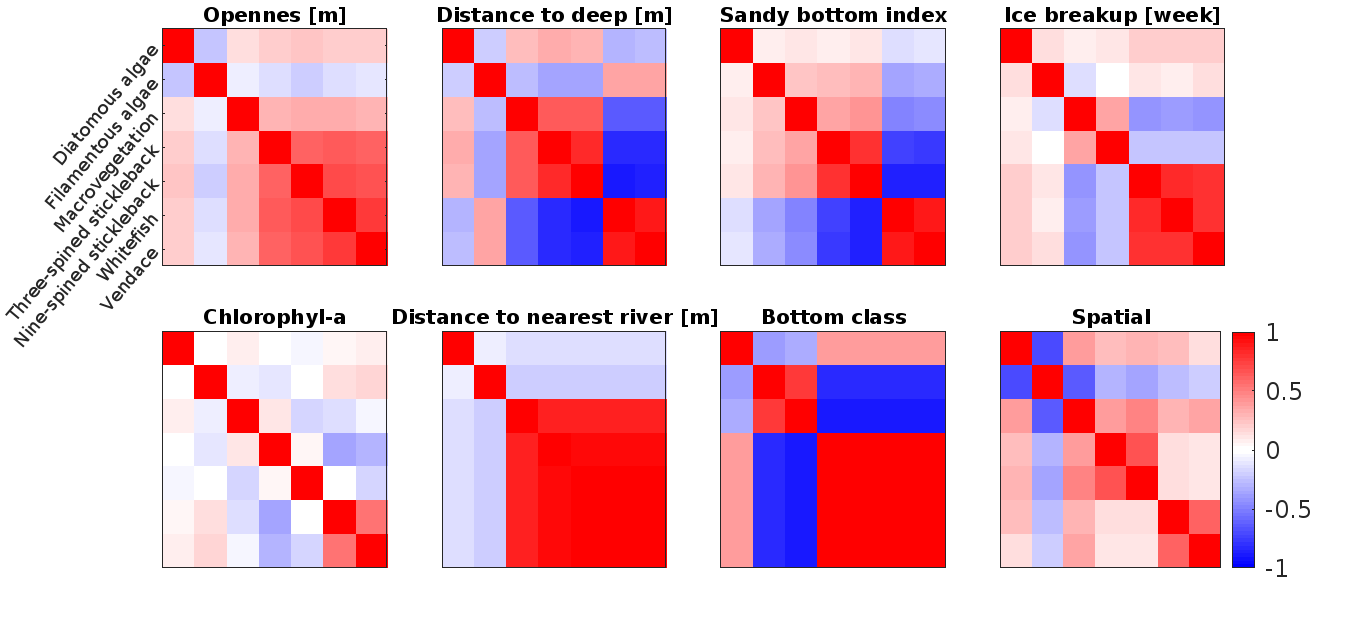}
\vspace{-0.8cm}
\caption{The maximum a posterior estimate of the correlation matrices in the multivariate additive GP model with Gaussian covariance function (model 3).}
\label{fig:correlations}
\vspace{-0.3cm}
\end{figure}

\subsection{Effects of environmental covariates}

\begin{figure}[!t]
\includegraphics[scale=0.54]{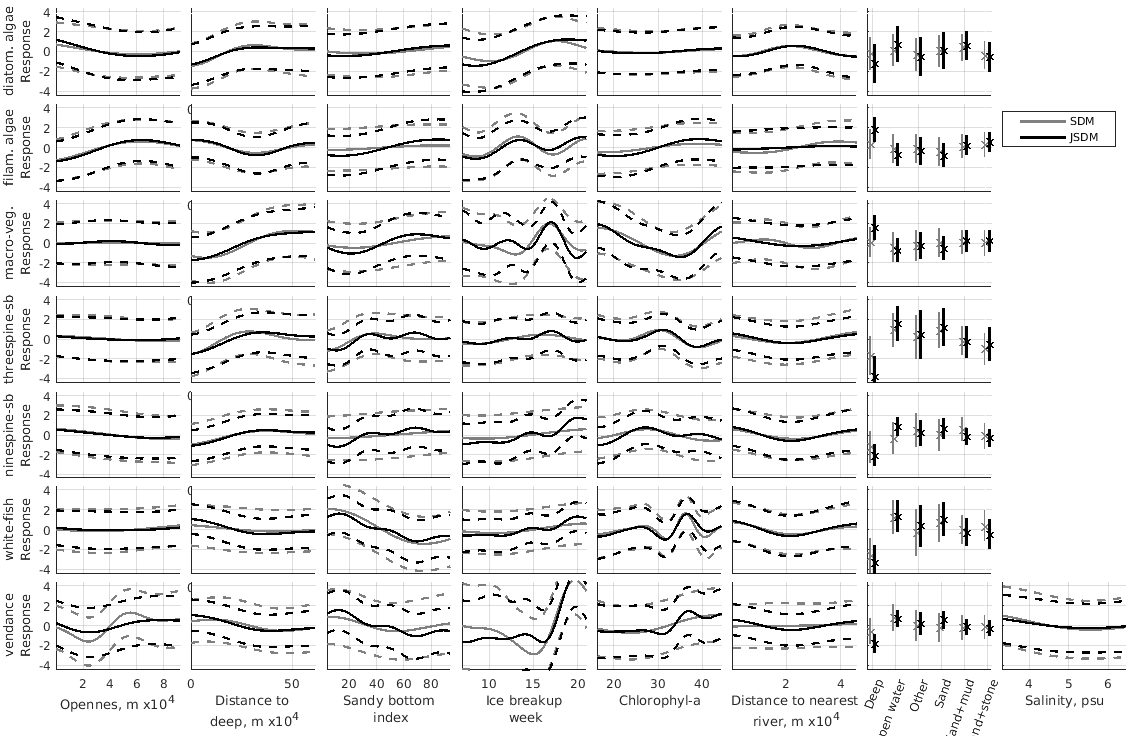}
\vspace{-0.7cm}
\caption{Marginal latent responses along covariates, $h_{j,r}$. Grey corresponds to single species GP model (model 1) and black to multivariate additive GP model (model 3). 
}\label{fig:responses}
\vspace{-0.5cm}
\end{figure}

The interspecific correlations (Figure~\ref{fig:correlations}) show that the responses to environment are similar among Coregonids and among sticklebacks whereas there are clear differences among these groups. These fish groups respond differently to sandy bottom and distance to deep. Moreover, there is strong positive spatial correlation among Coregonids and among sticklebacks but not between these groups. All species have negative spatial correlation with filamentous algae whereas there is either weak positive or negligible spatial correlation between fish species and macrovegetation and diatomous algae. 

Figure~\ref{fig:responses} shows the marginal effects of the predictor functions for both univariate (SSDM) and multivariate (JSDM) GP models calculated as described in section~\ref{sec:marginal_additive_affects}. 
In general, the responses in JSDM and SSDM are similar.
In most of the cases the uncertainty in response function is smaller (narrower credible regions) in JSDM than in SSDM although the differences are not large.
Most of the responses are smooth but show patterns that would be hard to capture with a quadratic function.
For example, three and nine spine stickleback, macrovegetation and filamentous algae show logistic style response to either distance to deep or openness.
Moreover, the response of vendace on ice break up week is first constant but has very steep increase after week 16. 
The responses to ice break up or chlorophyll-a show non-linear and non-quadratic responses also for white fish, macrovegetation and filamentous algae. 
The MAP estimates of other hyper-parameters are summarized in \ref{suppA}.

\subsection{Spatial predictions}

Figure~\ref{fig:prediction_map} shows the distribution maps as posterior median of intensity and expected count of individuals in replicate sampling produced by SSDM (model 1) and JSDM (model 3) for vendace. The distribution maps for other species together with maps on posterior predictive variance are shown in \ref{suppA}. 
In broad scale the posterior median of the intensity looks similar with SSDM and JSDM whereas SSDM predicts larger species counts than JSDM for all species throughout the study region and the uncertainty related to SSDM predictions is larger than that of JSDM predictions.
Both SSDM and JSDM predict that vendace is distributed mostly in the northern Gulf of Bothnia.
However, SSDM predicts somewhat higher median intensity than JSDM also in the southern Gulf of Bothnia.
In relation to median intensity similar pattern that JSDM predicts more restricted distribution range  than the single species model is seen also in the predictions concerning sticklebacks and whitefish. 
In case of diatomous and filamentous algae SSDM and JSDM predict the distribution pattern similarly whereas for macrovegetation JSDM predicts slightly larger distribution ranges. 
The posterior distributions of spatial lenght-scale parameters were concentrated near one kilometer or less (see table~5 in \ref{suppA}). Hence, the differences in distribution predictions cannot be explained by the spatial random effects over large areas but the spatial random effect explains local deviations from the predictions based on covariates.

\begin{figure}[!t]
\begin{center}
\includegraphics[scale=0.55]{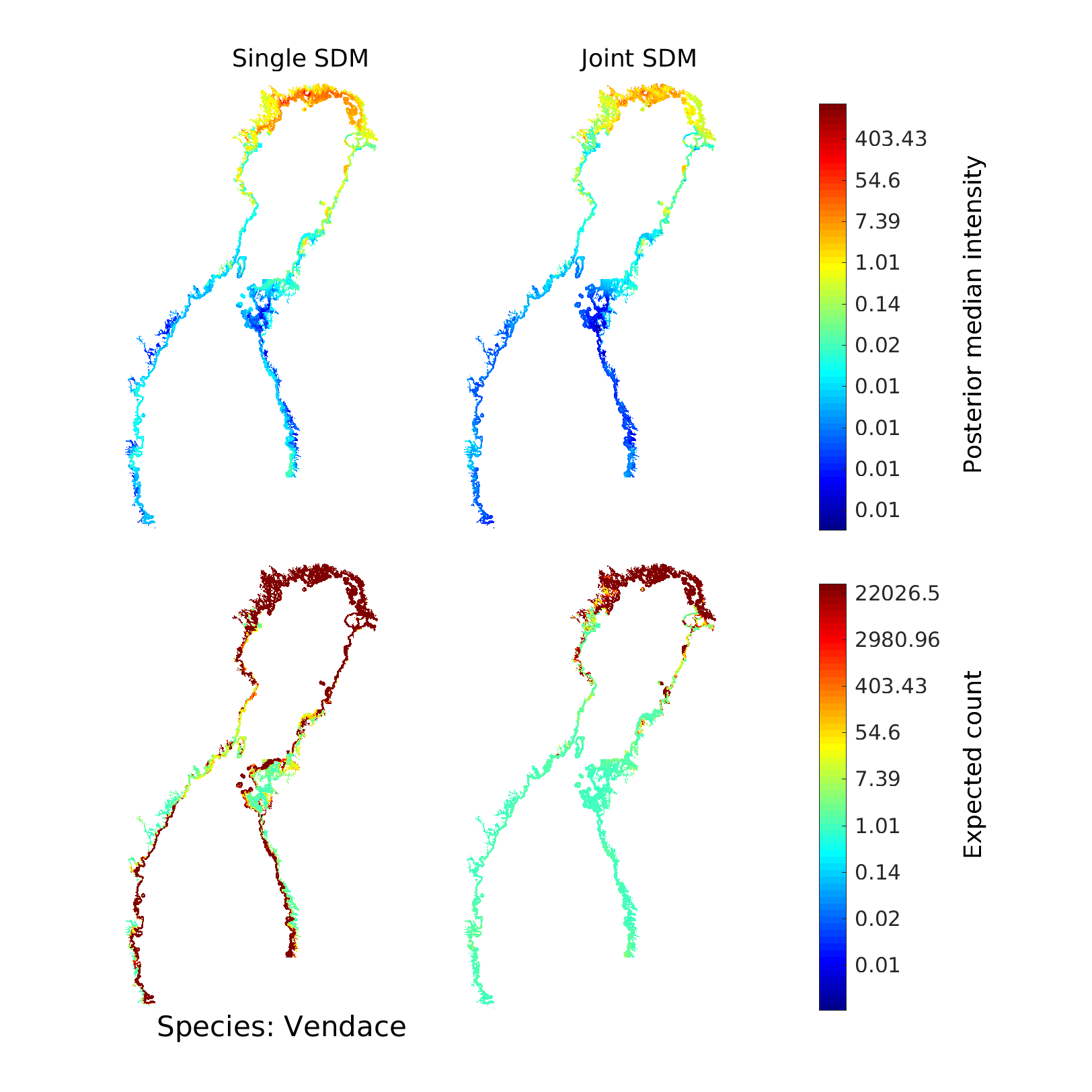}
\vspace{-0.6cm}
\caption{Posterior predictive median of intensity and the expected count of individuals in replicate sampling for vendace predicted by SSDM (model 1) and JSDM (model 3).}\label{fig:prediction_map}
\end{center}
\vspace{-0.5cm}
\end{figure}
%


\section{Discussion and concluding remarks}\label{sec:Discussion}

\subsection{Case study results}

The GP based SDMs had better predictive performance than the parametric SDMs and JSDMs had better predictive performance than the corresponding SSDMs. 
The differences between SSDM and JSDM are most apparent in the distribution maps and predictor functions.
The JSDM predicted in general smaller distribution ranges than SSDMs (macrovegetation being the only exception) and the posterior uncertainty in their predictions were smaller than in SSDMs.

When interpreting the results, it should be remembered that the sampling was targeted to larvae of sea-spawning Coregonids (whitefish and vendace) and planned to cover their plausible distribution range in terms of spatial (coastal areas of Gulf of Bothnia) and temporal extent (spring).
Other species were sampled alongside Coregonids and the sampling area covers only limited portion of their full distribution range in the spring. 
Moreover, the abundance and distribution of all the studied species vary annually. 
Fish change their distribution areas seasonally and their larval stages last only few weeks. 
Vegetation and algal cover vary due changes in temperature and ice effects.
For these reasons the results are most representative for larvae of whitefish and vendace. For other species the results describe their spring distribution in the shallow coastal regions only. 

Our results correspond rather well to the earlier knowledge on the studied species.
The responses to the environmental covariates and the interspecific correlations (Figures~\ref{fig:responses} and \ref{fig:correlations}) indicate that whitefish and vendace larvae are, in general, distributed in different areas than sticklebacks. 
Most of the literature on Baltic Sea sticklebacks focus on three-spined sticklebacks whereas the nine-spined stickleback is not well studied. 
In early spring, stickleback abundances have been found to be highest in sheltered archipelago areas, where part of the population overwinter. 
High abundance of sticklebacks are typically thought to indicate structural complexity on the bottom; such as the presence of stones and boulders as well as reeds and other macrovegetation that function as shelter and provide food \citep{Peltonenetal:2004,Siebenetal:2011}. 
On contrary, highest densities of whitefish and vendace larvae are observed in open sandy shores near deep areas and in structurally simple shores, without macrovegetation, boulders and stones \citep{Huddetal:1988,Leskelaetal:1991,lari:2013}.
The distribution of vendace is highly influenced by ice break up week and salinity so that long ice cover period and low salinity increase their abundance.
Ice winter is longer and salinity is lower in the northern than in the southern Gulf of Bothnia and thus it correlates positively to Coregonid presence \citep{vanh}.
In general, optimal habitats for Coregonid larvae are mainly located in the northernmost Gulf of Bothnia. 

Sticklebacks are known to prey mainly on mesozooplankton, but also on grazers \citep{Peltonenetal:2004,Siebenetal:2011}. 
Sticklebacks feed also on fish larvae if available \citep{Bystrometal:2015}, but there are no studies on potential predation risk to Coregonids. 
Based on our results, in the scale of coastal region of Gulf of Bothnia, sticklebacks are not thread to Coregonid larvae since their high density areas do not overlap. 
Moreover, there was no spatial correlation between sticklebacks and coregonids (Figure~\ref{fig:correlations}) which could indicate interspecific interaction of any kind.
The presence of sticklebacks has been connected to higher eutrophication status and stickleback reproduction and growth benefit from increasing temperature and eutrophication \citep{Leferbureetal:2011,Candolinetal:2008,Meieretal:2012} 
On contrary, these environmental characteristics are assumed to affect negatively the Coregonid reproduction \citep{Cingietal:2010,Muller:1992,lari:2013,vanh}. 
In our results whitefish and vendace larvae have positive response to Chlorophyl-a, which is a strong indicator for increasing eutrophication. 
However, in the scale of Gulf of Bothnia Chlorophyl-a concentration is typically higher near river mouths where salinity is lower and nutrient inflow high. 
Hence the result more likely reflects the high river influence through low salinity than preference for eutrophicated water.

The vegetation and algae distribution in our results reflect the nutrient status during winter and early spring as well as effect of ice cover and ice scraping in wind exposed shallow areas. 
Both the SSDMs and JSDMs indicate that filamentous algae occur in all coastal areas in high densities, except in some sheltered inner coastal areas. 
Macrovegetation and diatomous algae had highest presence probabilities at areas with lower filament presence probability. 
This pattern agrees well with the general understanding of these species groups. 
Filamentous algae are typically distributed in wind and wave exposed shores, that epiphytic diatoms and reeds that require shelter cannot tolerate. 
Filamentous algae can, however, grow over macrovegetation and it is possible that macrovegetation distribution is underestimated if filamentous algae growth over macrovegetation has hided macrovegetation from the sampling pictures. 
The higher nutrient levels in sheltered areas have been found to affect positively reed belt growth, and high abundances of macrovegetation species have been found from archipelago areas, lagoons, bays and river inlets \citep{Pitkanenetal:2013,Altartourietal:2014}. 
The JSDM model reflects these smaller scale occurrences better than SSDM. 
In our results, diatomous algae are more common in estuaries and northern parts of the study area (see \ref{suppA}). This is likely explained by longer ice winter towards northern Gulf of Bothnia and longer distance to deep water. 

\subsection{Multivariate additive Gaussian process}

Next we discuss some of the similarities and differences between our model and the existing JSDMs and highlight the novel methodological contributions in this paper. 

\subsubsection{Interspecific correlations}

From the predictive point of view, the interspecific correlations in predictive functions are attractive for many reasons. 
In many applications of SDMs the aim is to predict species distribution over regions that include locations spatially far from the data \citep{rec:2013} or to conduct scenario predictions related to, for example, climate change or land use \citep{guisan:2013}. 
In these applications, predictions are based on the responses to environmental covariates. 
The inclusion of interspecific dependence allows information flow between species which improves the estimates for predictive functions  especially for species with only scarce data \citep{thorson:2015,Ovaskainenetal:2017,Hui2013,clark+gelf:2014}.  

We used the marginally uniform priors of \citep{barnard,tokuda:2012} for the interspecific correlations. 
This is justified by prior ignorance on correlations. 
Prior information on interspecific correlations could, however, be added into model with, e.g., informative inverse Wishart prior. 
With many species inferring the full covariance matrix is hard in which case we could use spatially dependent latent factors which induce for the spatial random effects a covariance structure $\mathrm{Cov}\big(\epsilon_j(\s),\epsilon_j'(\s')\big)=\sum_{q=1}^M\lambda_{qj}\lambda_{qj'}k_q(\s,\s')$, where $k_q$ is the $q$'th spatial covariance function and $\lambda_{qj}$ the corresponding species specific factor loading.
Here, $M < J$ so that this covariance structure is a low rank representation of LMC \citep{lopez+2000,lopes+salazar+gamerman:2008}. 
Latent factor representation was first introduced to SDMs by \cite{thorson:2015} and it is used also in HMSC \citet{Ovaskainenetal:2017}. Recently \citet{Taylor-Rodriguez2016} proposed a clustering scheme where the species are clustered to less than $J$ factor loading vectors. 

Interspecific correlation between response functions has received less interest. Typically the response functions are assumed to be independent among species. Exception are species archetype models \citep{Dunstan2013} and the HMSC framework of \citet{Ovaskainenetal:2017}. In the latter, the response functions are defined as $h_j(\x_j) = \x_j\theta_{j\cdot}$ where $\theta=[\theta_{1\cdot}^T,\dots,\theta_{J\cdot}^T]^T$ is a $J\times p$ matrix of regression weights with hierarchical prior. 
The species specific weights are given Gaussian prior $\theta_{j\cdot}\sim N(\mu_{j\cdot},V_j)$ where $\mu_{j\cdot}$ is the expected response of a species that can be common for all species, common within groups of species or modelled through species traits $\mu_{jr}=\tau^T_{j}\gamma_{r}$ where $\tau_{j}$ is a vector of trait values of species $j$ and $\gamma_{r}\sim N(0,V_{\gamma})$ are the effects of traits to response along covariate $r$. 
With $V=\sigma^2I_{J\times J}$ and common prior mean $\mu_{j\cdot}=\mu\sim N(0,\sigma_{\mu}^2)$ for all $j\in 1,\dots,J$, the induced covariance between species specific additive response functions is 
\begin{align*}
\cov\left(h_{j,r}(x_r),h_{j',r}(x'_r)\right) = & \mathbb{E}[\cov\left(x_r\theta_{jr},x'_r\theta_{j'r}\right)] + \cov\left(\mathbb{E}[x_r\theta_{jr}], \mathbb{E}[x'_r\theta_{j'r}] \right)\\
=&(\sigma_{\mu}^2+\delta_j(j')\sigma^2)x_rx'_r
\end{align*}  
and with trait dependent prior mean $\cov\left(h_{j,r}(x_r),h_{j',r}(x'_r)\right) = (\tau_j^TV_{\gamma}\tau_{j'}+\delta_j(j')\sigma^2)x_rx'_r$. 
Hence, these alternatives have the same covariance structure as an additive multivariate GP \eqref{gp:dpd_3} with $k_{h_j,r}=x_rx'_r$ for all $j=1,\dots,J$, and interspecific covariances $\Sigma_{h_r}=\sigma_{\mu}^2+\delta_j(j')\sigma^2$ and $\Sigma_{h_r}=\tau_j^TV_{\gamma}\tau_{j'}+\delta_j(j')\sigma^2$. 
Hence, the hierarchical prior structure of HMSC could easily be extended to semiparametric models as well if we had trait information from our species. 
The benefit would be that these hierarchical model structures contain ecologically relevant prior information and, hence, using the induced interspecific covariance structure in the additive multivariate GP model would make it ecologically more realistic and potentially improve its predictive performance.
The HMSC framework has many other features as well, such as phylogenetic relationships \citet{Ovaskainenetal:2017}, which could in principle be incorporated with our approach as well.
The structure of interspecific correlations is likely to be especially important in scenario based predictive analyses, such as climate change predictions. 

Since our model does not make any mechanistic assumptions on species interdependencies the results concerning interspecific correlations have to be interpreted with care. 
For example, a positive correlation between spatial random effects could arize due mutualism, predator prey association or similar response to unobserved environmental covariates. 
Hence, interpreting these correlations should be done in light of more general ecological knowledge on the species. 
This is, however, a common challenge with all current JSDMs \citep{thorson:2015,Dunstan2013,Ovaskainenetal:2017,Taylor-Rodriguez2016}. 
Moreover, the property of our model, as well as most other JSDM, is that if the training data shows positive or negative spatial correlation between two species this positive correlation is assumed to hold everywhere in the study domain. 
In many cases this might be an unrealistic assumption for which reason one interesting future development would be to extend our models to allow species-to-species associations to depend on the environmental context\citep{fox:2015,tikhonov:2017}.
However, the colocalization patterns are explained by both environmental covariates and spatial random effects and regional differences in colocalization patterns can be explained by different environmental covariates in different regions. 

The interspecific correlation in our approach are modelled on the latent variable level and, hence, are not directly comparable to interspecific correlations in data.
Some authors have argued that this is a conceptual weakness of latent variable models since it makes interpretating their results hard and thus less transparent. 
An alternative would be to model the correlations in the observation error model \citep[see, e.g.,][]{otso:2010,laura+reid:2014}.
\citet{clark:2016} proposed generalized joint attribute model (GJAM) to fix this interpretation challenge by aiming to model the correlations between species on the data scale. 
However, for example, in case of presence absence observations GJAM corresponds to multivariate probit model that is the marginal likelihood of our model in case of Bernoulli observation model. 
It is also unclear how to disentangle the process underlying the species occurrence and abundance from the observation process leading to the actual data in GJAM approach.
Hence, despite the challenge of interpretating the correlations, we prefer the hierarchical latent variable modeling since it provides a coherent approach to separate these two processes.

\subsubsection{Response functions}

The response functions along environmental covariates provide basis to understand how environment affects species distribution and to predict the species distribution in new areas. 
Linear and other parametric models are efficient in finding overall trends in responses but have trouble in locating abrupt changes and change points due, for example, physical tolerance limits. 
Such limits, for example, along temperature and salinity are typical for large variety of taxa in aquatic domains \citep{MacKenzieetal:2007,Kottaetal:2019}. With SSDMs \citet{vanh}, \citet{Shelton2014}, \citet{golding:2016} and \citet{meri:2016} have demonstrated the benefits of semiparametric models in such situations. 
The GP approach for modeling environmental responses is similar to generalized additive models \citep{gui+ed:2002} but the latter have not been implemented as JSDMs.
The flexibility of semiparametric models provides also challenges compared to the more restricted parametric models. 
The prior for covariance function parameters has to be set with care \citep{golding:2016} and we may also need to pose monotonicity constraints to the response functions \cite{Kottaetal:2019} in order to prevent overfitting.
Inferring the responses reliably requires typically more data compared to parametric models. 
The multivariate additive GP helps in these challenges as well since the interspecific correlations increase the effectively amount of data to be used to infer each response function and hence decreases the uncertainty in them.

Additive multivariate GP framework opens also new questions related to the choice of covariance functions and interspecific correlations. 
A typical choice for general purpose covariance function is a radial basis function.
However, predictions using these covariance functions revert to prior predictive distributions when predicting beyond covariate range covered by data.
 Hence, in extrapolation tasks stationary covariance functions may not be the optimal choice \citep{vanh,meri:2016} and combining the multivariate GP models with functional constraints \citep[e.g.][]{Kottaetal:2019} could improve their predictive performance further.
 
\subsubsection{Spatial random effects}

Our model comparison results show clearly that including spatial random effects into model improves their predictive performance in both inter- and extrapolation. 
This result is well in line with earlier works demonstrating that  environmental conditions alone may not sufficiently explain species distribution and spatial random effects improve their predictive performance \citep{Latimer+etal:2006,vanh,clark+gelf:2014,thorson:2015,meri:2016}. 
This is reasonable, since the distribution of species is shaped by the interplay of environmental covariates, stochastic processes and species interaction \citep{Ovaskainenetal:2017}.  
Hence, a justified model should account for random processes and as demonstrated also by our results also to species interactions in these processes. 
However, adding spatial random effects into model may lead to problems in model identifibiality which we discuss in the next section.

\subsubsection{Posterior inference and predictive performance}

The JSDM built with multivariate additive GP had clearly the best overall predictive performance. 
Moreover, the uncertainty in the predictions by JSDMs were smaller than in the SSDMs which, together with the best log predictive density performance, indicates that the predictions were also more accurate. 
These findings have important implications to practical use of SDMs for management and other purposes. 
For example \citet{meri:2016} used SSDMs to classify coastal areas of Finland to unsuitable, suitable and highly productive spawning regions for four commercially important fish species. 
They based their estimates on the expected numbers of larvae in the coastal region which, as shown by Figure~\ref{fig:prediction_map} is highly sensitive to the uncertainty of the predictions. 
Moreover, by using SSDMs we can overestimate the total biomass of all species \citep{clark+gelf:2014}.

An inherent challenge with the type of models considered here is the model identifibiality. 
Spatial random effects are known to affect the fixed effects estimates \citep{Hodges2010} and in some cases the spatial random effect can actually capture variability otherwise explainable with fixed effects \citep{Paciorek2010,Bose2018a}.  
Moreover, with uniform (uninformative) priors the length-scale and variance hyperparameters of Mat{\'e}rn class of covariance functions are non-identifiable \citep{Zhang:2004} in general. 
In order to alleviate these identifibiality challenges we used weakly informative priors for the variance and length-scale parameters defined through the principles proposed by \cite{gelman:2006}, \cite{Simpson+etal:2014} and \cite{arne:2018}. 
Our priors for the covariance function parameters favor small variance and large length-scale parameter values. 
The former penalizes for large magnitudes and the latter penalizes for wiggly responses along covariates or space. 
In case of covariate responses this is ecologically justified since according to ecological niche theory species' responses to environmental covariates are typically either monotonic or have only one mode \citep{Ovaskainenetal:2016}.
Moreover, \cite{arne:2018} show that these type of priors work well for spatial random effects with Mat{\'e}rn covariance functions in general.
Favoring longer spatial length-scales is justified also from the model identifibiality point of view. 
Identifiability between fixed and spatial random effects is of less concern if spatial random effect varies with larger spatial range than the environmental covariates \citep[][Section~3]{Paciorek2010}.
Since the environmental covariates in Finnish coastal region vary with relatively small spatial range, our weakly informative prior effectively favors models that explain this variation with covariate responses instead of with spatial random effects.
Our model comparison suggests that the models with both covariate effects and spatial random effects lead to most reliable inference on environmental responses and spatial random effects.
A model's extrapolation performance is governed mainly by the environmental covariates so better extrapolation performance indicates more reliable response functions. 
Similarly more accurate interpolation performance indicates more reliable joint effect of environmental covariates and spatial random effect.

An evident challenge with multivariate additive GP is the computation. 
The core element in the multivariate additive GP is the covariance matrix induced by the GP components. 
With many species and sampling sites the covariance matrix can become so large that it makes the implementation infeasible. 
Here we used Laplace method to speed up the computation by decreasing the number of costly posterior density calculations compared to full MCMC.
However, in order to scale up the methods for large data sets involving thousands of sampling sites, the model would need to be implemented even more efficiently.
One option to reduce the computational time could be to exploit the property that linear LMC can be parameterized through species specific conditional distributions \citep{Gelfand+Schmidth+Banerjee+Sirmans:2004}.
We could also implement multivariate GPs with sparse GP approximations \citep{Vanhatalo+Pietilainen+Vehtari:2010, Alvarez+Luengo+Titsias+Lawrence:2010} and replace the full rank LMC model with latent factor model~\citep{thorson:2015,Ovaskainenetal:2017}. 
 However, we leave these considerations for future.

\subsection{Summary}

Our JSDM based on multivariate additive GP combines the key ideas from semi-parametric SSDMs and state-of-the-art parametric JSDMs. In our case study, it showed superior predictive performance compared to existing GP based SSDMs and parametric SSDMs and JSDMs in both interpolation and extrapolation.
Hence, the multivariate additive GP can be seen as the first step towards integration of the semi-parametric SSDMs and hierarchical JSDMs. 
We propose also an efficient approach for inference by utilizing Laplace approximation and gradient based optimization for hyperparameters.
The multivariate additive GP model is not restricted only to species distribution modeling but can be used in wide variety of other applications as well.


\begin{supplement} 
\sname{Supplementary material} \label{suppA}
\stitle{The supplementary material contains additional mathematical formulation of the methodology proposed in this paper and additional figures and tables for the case study analysis.}
\slink[url]{http://www.some-url-address.org/dowload/0000.zip}
\end{supplement}

%

\begin{acknowledgement}
This work has bee funded by the Academy of Finland (grant 317255) and Research Funds of the University of Helsinki (decision No. 465/51/2014).
\end{acknowledgement}

\bibliographystyle{ba}
\bibliography{refs}

\end{document}


\begin{frontmatter}
\title{Supplementary Material: Additive multivariate Gaussian process for joint species distribution modeling with heterogeneous data}
\runtitle{Supplement: JSDMs with additive multivariate Gaussian processes}

\begin{aug}
\author{\fnms{Jarno} \snm{Vanhatalo}\thanksref{addr1}\ead[label=e1]{jarno.vanhatalo@helsinki.fi}},
\author{\fnms{Marcelo} \snm{Hartmann}\thanksref{addr2}\ead[label=e2]{marcelo.hartmann@helsinki.fi}}
\and
\author{\fnms{Lari} \snm{Veneranta}\thanksref{addr3}}
\ead[label=e3]{third@somewhere.com}
\runauthor{Vanhatalo, Hartmann and Veneranta}

\address[addr1]{Department of Mathematics and Statistics and Organismal and Evolutionary Biology Research Program, University of Helsinki, 
    \printead{e1} 
}

\address[addr2]{Department of Mathematics and Statistics, University of Helsinki,
    \printead{e2}
}

\address[addr3]{Natural Resources Institute Finland, Finland}
\end{aug}

\end{frontmatter}

\section{The posterior expectation and variance for new observations}\label{app:mean+var}

First we approximate the logistic function with $p(f)\approx \tilde{p}(f) = a \Phi(f/v_1) + (1-a) \Phi(f/v_2)$ where $a$ $\in$ $(0, 1)$ and $v_1, v_2$ $\in$ $(0, \infty)$ are parameters of the approximation \citep[see][]{demidenko:2004}. This approximation is used since it has small error bound ($\leq |10^{-4}|$, \cite{demidenko:2004}) and can considerably speed-up marginalization over $f$ for large datasets. Using Lemma 2 in the Section \ref{app:GaussInt}, we obtain, 
\begin{align} \label{eq:predBinE}
\Ex[Y_j(\x_*, \s_*)|\y, \bbeta, \Lambda] &= z_{j, *} a\Phi\Big(\tfrac{\mu_{j, *}}{\sqrt{\Sigma_{j, *} + v^2_1}}\Big) +  z_{j, *} (1 - a)\Phi\Big(\tfrac{\mu_{j, *}}{\sqrt{\Sigma_{j, *} + v^2_2}}\Big) \\
\Vx[Y_j(\x_*, \s_*)|\y, \bbeta, \Lambda] &= \Ex[Y_j(\x_*, \s_*)|\y, \bbeta, \Lambda] - \Ex[Y_j(\x_*, \s_*)|\y, \bbeta, \Lambda]^2 \nonumber \\
&\phantom{=} + (z_{j, *}^2 - z_{j, *}) \mathbb{E}[\tilde{p}^2(f_*)] \nonumber 
\end{align}
where, $\Phi(\cdot)$ is the cumulative distribution of the standard Gaussian random variable and 
\begin{equation}
\mathbb{E}[\tilde{p}^2(f_*)] = a^2 F_2(\bmu_{j, *}|V_{1, 1}) + (1-a)^2 F_2(\bmu_{j, *}|V_{2, 2}) + 2a(1-a)F_2(\bmu_{j, *}|V_{1, 2})
\end{equation}
$F_2(\bmu_{j, *}|V_{m, n})$ is the zero-mean 2-dimensional Gaussian cumulative distribution function evaluated at $\bmu_{j, *}$ $=$ $[\mu_{j, *}$ $ \mu_{j, *}]^T$ with covariance matrix given by
\begin{equation}
V_{m, n} = 
\begin{bmatrix} \Sigma_{j, *} + v^2_m & \Sigma_{j, *} \\
\Sigma_{j, *} & \Sigma_{j, *} + v^2_n
\end{bmatrix}.
\end{equation}

In the case of Negative-Binomial model, we use the moment generating function of a Gaussian random variable to obtain the unconditional expectation of a future outcome w.r.t to latent variable, i.e.
\begin{align} \label{eq:predNegBinE}
\Ex[Y_{j}(\x_*, \s_*)|\y, \bbeta, \Lambda] &= z_{j, *} e^{\mu_{j, *} + \Sigma_{j, *}/2} \\
\Vx[Y_{j}(\x_*, \s_*)|\y, \bbeta, \Lambda] & = z_{j, *} e^{\mu_{j, *} + \Sigma_{j, *}/2} + z_{j, *}^2 e^{2\mu_{j, *} + \Sigma_{j,*}} \left[ e^{\Sigma_{j, *}/2}(\hat{r}_{j} + 1)/\hat{r}_{j} - 1 \right]. \nonumber
\end{align}

\section{Conditional predictions}\label{app:cond_pred}

Recall that, in the main paper, the distribution of $\Y_{\mathcal{J}_1}(\x_*, \s_*)|\y_{\mathcal{J}_2}$ conditioned on hyperparameters (they were ommited from the notation for simplicity) is given by
\begin{align} \label{eq:condpred2}
\pi(\y_{\mathcal{J}_1, *}|\y_{\mathcal{J}_2}) 
&= \displaystyle\int
\pi(\y_{\mathcal{J}_1, *}| \y_{\mathcal{J}_2}, \f_{\mathcal{J}_1, *})\pi(\y_{\mathcal{J}_2}, \f_{\mathcal{J}_1, *}) \mathrm{d}\hspace{-0.04cm}\f_{\mathcal{J}_1, *}/ \pi(\y_{\mathcal{J}_2})
\end{align}
The second term in the numerator of the integrand of \eqref{eq:condpred2} is the same as,
\begin{align} \label{eq:condpred3}
\pi(\y_{\mathcal{J}_2}, \f_{\mathcal{J}_1, *}) 
&= \displaystyle\int
\pi(\f_{\mathcal{J}_1, *}| \f_{\mathcal{J}_2}, \y_{\mathcal{J}_2})\pi(\f_{\mathcal{J}_2}|\y_{\mathcal{J}_2})\pi(\y_{\mathcal{J}_2})\mathrm{d}\hspace{-0.04cm}\f_{\mathcal{J}_2}.
\end{align}
Now, $\f_{\mathcal{J}_1, *}|\f_{\mathcal{J}_2}, \y_{\mathcal{J}_2}$ also only depends on $\f_{\mathcal{J}_2}$. Plugging \eqref{eq:condpred3} into \eqref{eq:condpred2} we get
\begin{align} \label{eq:condpred5}
\pi(\y_{\mathcal{J}_1, *}|\y_{\mathcal{J}_2}) &= \displaystyle\int
\pi(\y_{\mathcal{J}_1, *}|\f_{\mathcal{J}_1, *}) \pi(\f_{\mathcal{J}_1, *}| \f_{\mathcal{J}_2})\pi(\f_{\mathcal{J}_2}|\y_{\mathcal{J}_2})\mathrm{d}\hspace{-0.04cm}\f_{\mathcal{J}_2}\mathrm{d}\hspace{-0.04cm}\f_{\mathcal{J}_1, *}
\end{align}
where last two terms in \eqref{eq:condpred5} are Gaussians. The first one is the conditional Gaussian density function w.r.t. the predictor function values $\f_{\mathcal{J}_2}$, that is, $\pi(\f_{\mathcal{J}_1, *}| \f_{\mathcal{J}_2})$ $=$ $\mathcal{N}(\f_{\mathcal{J}_1, *}|C_{\mathcal{J}_1, \mathcal{J}_2*}C_{\mathcal{J}_2, \mathcal{J}_2}^{-1}\f_{\mathcal{J}_2}, C_{\mathcal{J}_1{*, *}} - C_{\mathcal{J}_1, \mathcal{J}_2*} C_{\mathcal{J}_2, \mathcal{J}_2}^{-1} C_{\mathcal{J}_1, \mathcal{J}_2*}^T)$ and the second one is the Laplace approximation using the \emph{scenario species data} related to the set $\mathcal{J}_2$, that is, $\pi(\f_{\mathcal{J}_2}|\y_{\mathcal{J}_2})$ $\approx$ $\mathcal{N}\big(\hspace{-0.05cm} \f_{\mathcal{J}_2}| C_{\mathcal{J}_2, \mathcal{J}_2} \nabla \log \pi(\y_{\mathcal{J}_2}|\hat{\f}_{\mathcal{J}_2}), (C_{\mathcal{J}_2, \mathcal{J}_2}^{-1} + \nW_{\mathcal{J}_2})^{-1}\big)$. Plugging this approximate density function in the equation \eqref{eq:condpred5} gives 
\begin{align} \label{eq:condpred6}
\pi(\y_{\mathcal{J}_1, *}|\y_{\mathcal{J}_2}) &= \displaystyle\int
\pi(\y_{\mathcal{J}_1, *}|\f_{\mathcal{J}_1, *})\mathcal{N}(\f_{\mathcal{J}_1, *}|\mu_{|\mathcal{J}_2}, \Sigma_{|\mathcal{J}_2})\mathrm{d}\hspace{-0.04cm}\f_{\mathcal{J}_1, *}
\end{align}
with
\begin{align} \label{eq:condpred7}
\mu_{|\mathcal{J}_2} &= C_{\mathcal{J}_1, \mathcal{J}_2*} \nabla \log \pi(\y_{\mathcal{J}_2}|\hat{\f}_{\mathcal{J}_2}) \nonumber \\
\Sigma_{|\mathcal{J}_2} &= C_{\mathcal{J}_1{*, *}} - C_{\mathcal{J}_1, \mathcal{J}_2*} (C_{\mathcal{J}_2, \mathcal{J}_2} + \nW_{\mathcal{J}_2}^{-1})^{-1} C_{\mathcal{J}_1, \mathcal{J}_2*}^T
\end{align}
where $C_{\mathcal{J}_1, \mathcal{J}_2*}$ is the covariance between species in the set $\mathcal{J}_1$ at the point $(\x_*, \s_*)$ and the species in the set $\mathcal{J}_2$ in their respective covariates and spatial locations. $\nabla \log \pi(\y_{\mathcal{J}_2}|\hat{\f}_{\mathcal{J}_2})$ correspond to the gradient of the log-likelihood function for species related to the set $\mathcal{J}_2$. $C_{\mathcal{J}_1{*, *}}$ is the covariance matrix of $\f_{\mathcal{J}_1, *}$ at $(\x_*, \s_*)$. $C_{\mathcal{J}_2, \mathcal{J}_2}$ is the covariance matrix of $\f_{\mathcal{J}_2}$ and $\nW_{\mathcal{J}_2}$ is the negative Hessian matrix of the negative logarithm of the likelihood functions related to species in the set $\mathcal{J}_2$. Now, to calculate the predictive mean vector and predictive covariance matrix we follow the steps used in deriving unconditional expectation and unconditional variances as in the main paper. However we exclude them for brevity.


\section{Unconstrained parametrization of covariance matrices} \label{app:derLMC}

Let the matrix $\Sigma$ be a covariance matrix of dimension $J$. In terms of variances and correlations, we rewrite $\Sigma = \diag(\sigma^2_1, \ldots, \sigma^2_J)^{\frac{1}{2}}\Rho\diag(\sigma^2_1, \ldots, \sigma^2_J)^{\frac{1}{2}}$, where $\sigma^2_j$, $j$ $=$ $1, \ldots, J$ are variances and these are transformed as $\sigma^2_j$ $=$ $\exp(\delta_j)$. The correlation matrix $\Rho$ is written in terms of its Cholesky decomposition, that is $\Rho = L L^T$, where $L$ is the lower triangular Cholesky decomposition. In our case, the upper triangular Cholesk decomposition is given by 
\begin{equation} \label{eq:chols}
L^T = 
\setlength\arraycolsep{4pt}
\begin{bmatrix*}[r]
1 & z_{1, 2} & z_{1, 3} & \cdots & z_{1, J} \\[0.3cm] 
0 & \prod_{i=1}^1(1-z_{i, 2}^2)^\frac{1}{2} & z_{2, 3}\prod_{i=1}^1(1-z_{i, 3}^2)^\frac{1}{2} & \cdots & z_{2, J}\prod_{i=1}^1(1-z_{i, J}^2)^\frac{1}{2} \\[0.3cm]
\vdots & \vdots & \vdots & \ddots & \vdots \\[0.3cm]
0 & 0 & 0 & \cdots & \prod_{i=1}^{J-1}(1-z_{i, J}^2)^\frac{1}{2}
\end{bmatrix*} 
\end{equation}
where each $z_{i, i'}$ $\in$ $(-1, 1)$. Now, since each $z_{i, i'}$ can freely vary in the interval $(-1, 1)$ without violate the positive definiteness property of $\mathcal{P}$ \citep[see][]{kuro:2003, lkj:2009}, we map each $z_{i, i'}$ to the real line as $z_{i, i'}$ $=$ $2/[1 + \exp(-a \delta_{i, i'})] - 1$, where $\delta_{i, i'}$ $\in$ $\mathbb{R}$.

The derivative of the $\log \pi(\Rho)$ (recall the prior for the correlation matrix in the main paper) w.r.t. each $\delta_{i, i'}$ is given by
\begin{equation}
\dfrac{\partial }{\partial \delta_{i, i'}} \log \pi(\Rho|v) = \sum_{j = 1}^{J-1}  \sum_{j' = j + 1}^{J} \dfrac{\partial }{\partial \rho_{j, j'}} \log \pi(\Rho|v) \dfrac{\partial }{\partial \delta_{i, i'}}  \rho_{j, j'}
\end{equation}
and the partial derivatives of $\log \pi(\Rho|v)$ w.r.t to $\rho_{j, j'}$ are obtained as,
\begin{equation}
\dfrac{\partial}{\partial \rho_{j, j'}}  \log \pi(\Rho|v) = \big((v-1)(J-1)-1\big) \lbrace \Rho^{-1} \rbrace_{j, j'}  - v \sum\limits_{\substack{\ \ r = 1 \ : \\[0.12cm] r \notin \lbrace j, j' \rbrace}}^{J} \lbrace \Rho_{rr}^{-1} \rbrace_{c, c'}
\end{equation}
where $c = j - 1$ and $c' = j'$ if $r < j$ and, $j' = 1$ or $r > j'$. If $r < j$ and $r < j'$ then $c = j - 1$ and $c' = j' - 1$. $\lbrace \Rho^{-1} \rbrace_{j, j'}$ is the entry $(j, j')$ of the matrix $\Rho^{-1}$ and $\lbrace \Rho_{rr}^{-1} \rbrace_{c, c'}$ is the entry $(c, c')$ of the inverse matrix $\Rho_{rr}^{-1}$, where $\Rho_{rr}$ is a matrix obtained by removing the $r$'th row and column of $\Rho$. The partial derivatives of each $\rho_{j, j'}$ w.r.t to $\delta_{i, i'}$ are obtained from $\partial \hspace{-0.05cm} \Rho \hspace{-0.05cm} / \partial \delta_{i, i'}$ $=$ $(\partial L/\partial\delta_{i, i'}) L^T + L (\partial L/\partial\delta_{i, i'})^T$.

Lastly, we find the derivatives of the logarithm of the absolute value of the Jacobian determinant w.r.t each $\delta_{i, i'}$ (see equation (11) in \citealt{lkj:2009}).
\begin{eqnarray}
\dfrac{\partial}{\partial \delta_{i, i'}} \log \left| \dfrac{\partial (\rho_{1, 2}, \ldots, \rho_{J-1, J})}{\partial(\delta_{1, 2}, \ldots, \delta_{J-1, J})} \right| =& 
\dfrac{\partial}{\partial \delta_{i, i'}} \log \displaystyle\prod_{j = 1}^{J-1} \prod_{j' = j + 1}^{J} (1-z_{j, j'}^2)^{(J - j - 1)/2} \dfrac{\mathrm{d} z_{j, j'}}{\mathrm{d} \delta_{j, j'}}
\nonumber \\
=& - (J-i-1) \dfrac{z_{i, i'}}{1 - z_{i, i'}^2}\dfrac{\mathrm{d} z_{i, i'}}{\mathrm{d} \delta_{i, i'}} + \dfrac{\mathrm{d}^2 z_{i, i'}/\mathrm{d} \delta_{i, i'}^2}{\mathrm{d} z_{i, i'}/\mathrm{d} \delta_{i, i'}}
\end{eqnarray}

\section{Gaussian integrals} \label{app:GaussInt}

\begin{lem} \label{lemma:2} Let's denote by $\Phi(\cdot)$ the standard-Gaussian cumulative distribution function and $\mathcal{N}(\cdot|\mu, \sigma^2)$ the Gaussian density function with parameters $(\mu, \sigma^2) \in \mathbb{R} \times \mathbb{R}_+$. Then the following holds
\begin{equation} \label{GaussInt}
\int
\limits_{-\infty}^{\infty}\mathcal{N}(x|\mu, \sigma^2) 
\prod\limits_{r = 1}^{N}\Phi\left(\dfrac{x-m_r}{v_r}\right) \mathrm{d}x = F_N\left(\bmu_N \vert \mathbf{m}_N, V_N \right) 
\end{equation}
where $F_N(\cdot|\mathbf{c}, \mathcal{C})$ is the $N$-dimensional Gaussian cumulative distribution function with parameters $(\mathbf{c}, \mathcal{C}) \in \mathbb{R}^N \times \mathcal{R}$, with $\mathcal{R}$ the space of positive-definite matrices (covariance matrices). Furthermore, $\mathbf{m}_N = [m_1 \cdots m_N]^T \in \mathbb{R}^N$, $\bmu_N = \mu \mathbf{1}_N \in \mathbb{R}^N$, $v_r > 0 \ \forall r$ and $V_N$ is a covariance matrix given by,
\begin{equation}
V_N = \begin{bmatrix}
v_1^2 + \sigma^2 & \ldots & \sigma^2 \\ 
\vdots & \ddots & \vdots \\ 
\sigma^2 & \ldots & v_N^2 + \sigma^2
\end{bmatrix}
\end{equation}
\end{lem}
\begin{proof}
To show \eqref{GaussInt}, start writing the left-hand side of the equation in full. Rewrite the integrand as the product of non-standard Gaussian density functions as well as the regions of integration, i.e., 
\begin{equation} \label{GaussInt1}
\int
\limits_{-\infty}^{\infty} 
\int\limits_{-\infty}^x \cdots \int\limits_{-\infty}^x \mathcal{N}(y_1|m_1, v_1^2)\hdots\mathcal{N}(y_N|m_N, v_N^2) \mathcal{N}(x|\mu, \sigma^2)\mathrm{d}y_1 \cdots \mathrm{d}y_N \mathrm{d}x.
\end{equation}
Rewrite again using the following transformation $[x, y_1, \cdots, y_N]^T =  [w + \mu, z_1 + w + m_1, \cdots, z_N + w + m_N]^T$ and note that $|\partial (x, y_1, \cdots, y_N)/\partial (w, z_1, \cdots, z_N)| = 1$. After changing variables, group the different terms in the exponentials together to have
\begin{equation} \label{GaussInt2}
\int \limits_{-\infty}^{\infty} 
\int\limits_{-\infty}^{\mu - m_N} \cdots \int\limits_{-\infty}^{\mu - m_1} \tfrac{1}{c} \exp \left\lbrace -\dfrac{1}{2} \left[ \sum_{r = 1}^N \tfrac{(z_r + w)^2}{v_r^2} + \tfrac{w^2}{\sigma^2} \right] \right\rbrace \mathrm{d}z_1 \ldots \mathrm{d}z_N \mathrm{d}w
\end{equation}
where $c = \sigma (2\pi)^{(N+1)/2} \prod_{r=1}^N v_r$. Now, the expression inside the squared bracket is a quadratic form which is written with the following matrix form,
\begin{align} \label{quadForm}
\sum_{r = 1}^N \tfrac{(z_r + w)^2}{v_r^2} + \tfrac{w^2}{\sigma^2} =& \ w^2 \left(\sum_{r = 1}^N \tfrac{1}{v^2_r} + \tfrac{1}{\sigma^2} \right) + w \sum_{r = 1}^N \tfrac{z_r}{v^2_r} + \sum_{r = 1}^N z_r \left( \tfrac{w}{v^2_r} + \tfrac{z_r}{v^2_r} \right) \nonumber \\[0.3cm]
= & \begin{bmatrix}
w \left(\sum_{r = 1}^N \tfrac{1}{v^2_r} + \tfrac{1}{\sigma^2} \right) + \sum_{r = 1}^N \tfrac{z_r}{v^2_r} \\[0.3cm]
\tfrac{w}{v^2_1} + \tfrac{z_1}{v^2_1} \\
\vdots \\
\tfrac{w}{v^2_N} + \tfrac{z_N}{v^2_N}
\end{bmatrix}^T
\begin{bmatrix}
w \\ 
z_1 \\ 
\vdots \\
z_N
\end{bmatrix} \nonumber \\[0.3cm]
= & \begin{bmatrix}
w \\ 
z_1 \\ 
\vdots \\ 
z_N
\end{bmatrix}^T  
\begin{bmatrix}
\sum_{r = 1}^N \tfrac{1}{v_r^2} + \tfrac{1}{\sigma^2} & \tfrac{1}{v_1^2} & \cdots & \tfrac{1}{v_N^2} \\ 
\tfrac{1}{v_1^2} & \tfrac{1}{v_1^2} & \cdots & 0 \\ 
\vdots & \vdots & \ddots & \vdots \\ 
\tfrac{1}{v_N^2} & 0 & \cdots & \tfrac{1}{v_N^2}
\end{bmatrix} 
\begin{bmatrix}
w \\ 
z_1 \\ 
\vdots \\ 
z_N
\end{bmatrix} 
\end{align}
therefore \eqref{GaussInt2} is the same as
\begin{equation} \label{GaussInt3}
\int
\limits_{-\infty}^{\infty} 
\int\limits_{-\infty}^{\mu - m_N} \cdots \int\limits_{-\infty}^{\mu - m_1} \tfrac{1}{c} \exp \left\lbrace  -\dfrac{1}{2} \begin{bmatrix}
w \\ 
z_1 \\ 
\vdots \\ 
z_N
\end{bmatrix}^T  
\begin{bmatrix}
\sum_{r = 1}^N \tfrac{1}{v_r^2} + \tfrac{1}{\sigma^2} & \tfrac{1}{v_1^2} & \cdots & \tfrac{1}{v_N^2} \\ 
\tfrac{1}{v_1^2} & \tfrac{1}{v_1^2} & \cdots & 0 \\ 
\vdots & \vdots & \ddots & \vdots \\ 
\tfrac{1}{v_N^2} & 0 & \cdots & \tfrac{1}{v_N^2}
\end{bmatrix} 
\begin{bmatrix}
w \\ 
z_1 \\ 
\vdots \\ 
z_N
\end{bmatrix} 
\right\rbrace \mathrm{d}\hspace{-0.04cm}\bz \mathrm{d}w \\[0.3cm]
\end{equation}
The integrand in \eqref{GaussInt3} has the full form of the multivariate Gaussian density. To identify this we need to find the closed-form covariance matrix from the precision matrix in \eqref{GaussInt3} and show that the determinant of the covariance matrix is given by $c^2/(2\pi)^{N+1}$. 
Write the precision matrix as block matrix such that $A = \sum_{r = 1}^N \tfrac{1}{v_r^2} + \tfrac{1}{\sigma^2}$, $B = \Big[\tfrac{1}{v_1^2} \cdots \tfrac{1}{v_N^2}\Big]$, $ C = B^T$ and $D = \diag\Big(\tfrac{1}{v_1^2}, \cdots, \tfrac{1}{v_N^2} \Big)$. 
Use the partitioned matrix inversion lemma \citep[][equation 17.44]{strang:1997}  to get the blocks, $(A - BD^{-1}C)^{-1} = \sigma^2$, $(BD^{-1}C - A)^{-1}BD^{-1} = -\sigma^2 [1 \cdots 1]$, $D^{-1}C(BD^{-1}C - A)^{-1} = -\sigma^2 [1 \cdots 1]^T$ and $D^{-1} + D^{-1}C(A - BD^{-1}C)^{-1}BD^{-1}$ where its main diagonal equals to $[v_1^2 + \sigma^2, \cdots, v_N^2 + \sigma^2]$ and all off-diagonal elements are given by $\sigma^2$. Put everything together to have the covariance matrix 
\begin{equation} \label{fullCov}
\begin{bmatrix}
\sigma^2 & -\sigma^2 & \cdots & -\sigma^2 \\ 
-\sigma^2 & v_1^2 + \sigma^2 & \cdots & \sigma^2 \\ 
-\sigma^2 & \vdots & \ddots & \vdots \\ 
-\sigma^2 & \sigma^2 & \cdots & v_N^2 + \sigma^2
\end{bmatrix} 
\end{equation}
whose determinant equals to $ 1/[\det(D)\det(A-BD^{-1}C)] = \sigma^2 \prod_{r = 1}^{N} v^2_r = c^2/(2\pi)^{N+1}$ by the partitioned matrix determinant lemma \citep[e.g.,][]{Rasmussen+Williams:2006}. Finally, in \eqref{GaussInt3}, interchange the order of integration with Fubini-Tonelli theorem \citep{folland:2013} and integrate w.r.t. $w$ to get
\begin{equation} \label{GaussInt4}
\int\limits_{-\infty}^{\mu - m_N} \cdots \int\limits_{-\infty}^{\mu - m_1}  \mathcal{N} \left(\begin{bmatrix}
z_1 \\ 
\vdots \\ 
z_N
\end{bmatrix} \Bigg| 
\begin{bmatrix}
0 \\
\vdots \\
0
\end{bmatrix}, 
\begin{bmatrix}
v_1^2 + \sigma^2 & \cdots & \sigma^2 \\ 
\vdots & \ddots & \vdots \\ 
\sigma^2 & \cdots & v_N^2 + \sigma^2
\end{bmatrix} \right) \mathrm{d}z_1 \cdots \mathrm{d}z_N
\end{equation}
that equals to
\begin{equation} \label{GaussInt5}
F_N \left(\begin{bmatrix}
\mu \\ 
\vdots \\ 
\mu
\end{bmatrix} \Bigg| 
\begin{bmatrix}
m_1 \\
\vdots \\
m_N
\end{bmatrix}, 
\begin{bmatrix}
v_1^2 + \sigma^2 & \cdots & \sigma^2 \\ 
\vdots & \ddots & \vdots \\ 
\sigma^2 & \cdots & v_N^2 + \sigma^2
\end{bmatrix} \right)
\end{equation}
and therefore the equality \eqref{GaussInt} holds. 
\end{proof}


\begin{lem} \label{lemma:3}
Let $\mathcal{N} (\cdot|\bmu_N, \Sigma)$ be the $N$-dimensional Gaussian density function with mean parameter $\bmu_N$ and covariance matrix $\Sigma$. Then the following holds true,
\begin{equation} \label{AnGaussInt}
\int\limits_{\mathbb{R}^N}
F_N(\x|\mathbf{m}_N, \mathrm{V})\mathcal{N}(\x|\bmu_N, \Sigma) \mathrm{d} \hspace{-0.04cm}\x = F_N\left(\bmu_N \vert \mathbf{m}_N, \mathrm{V} + \Sigma \right) 
\end{equation}
where $F_N(\cdot|\mathbf{c}, \mathcal{C})$ is the $N$-dimensional Gaussian cumulative distribution function with parameters $(\mathbf{c}, \mathcal{C})$. Furthermore, $\mathbf{m}_N = [m_1  \cdots  m_N]^T \in \mathbb{R}^N$, $\bmu_N = [\mu_1 \cdots \mu_N]^T \in \mathbb{R}^N$, $\mathrm{V}$ and $\Sigma$ are covariance matrices.
\end{lem}
\begin{proof}
Let's rewrite the left-hand side of \eqref{AnGaussInt} in full,
\begin{equation} 
\int\limits_{\mathbb{R}^N}
\int\limits_{-\infty}^{x_n} \cdots \int\limits_{-\infty}^{x_1}\mathcal{N}(\y|\mathbf{m}, \mathrm{V})\mathcal{N}(\x|\bmu_N, \Sigma)\mathrm{d}\mathbf{y} \mathrm{d}\mathbf{x}
\end{equation}
where $\y = [y_1, \cdots, y_n]^T$. Let's use the following transformation $[x_1, \cdots, x_N, y_1, \cdots, y_N]^T$ $=$ $[w_1 + \mu_1, \cdots, w_N + \mu_N, z_1 + w_1 + m_1, \cdots, z_N + w_N + m_N]^T$. The Jacobian of this transformation simplifies to $|\partial(x_1, \cdots, x_N, y_1, \cdots, y_N)/$ $\partial (w_1, \cdots, w_N, z_1, \cdots, z_N)| = 1$. Rewrite the above equation to get, 
\begin{equation} \label{AnGaussInt1}
\int\limits_{\mathbb{R}^N}
\int\limits_{-\infty}^{\mu_N - m_N} \cdots \int\limits_{-\infty}^{\mu_1 - m_1}\mathcal{N}(\mathbf{w}|-\mathbf{z}, \mathrm{V})\mathcal{N}(\mathbf{w}|\0, \Sigma)\mathrm{d}\mathbf{z} \mathrm{d}\mathbf{w}
\end{equation}
where $\mathbf{z} = [z_1 \cdots z_N]^T$ and $\mathbf{w} = [w_1 \cdots w_N]^T$. Note that the product of two multivariate Gaussian density functions is another unnormalized multivariate Gaussian \citep[see, e.g.,][Appendix A]{Rasmussen+Williams:2006}. Therefore we write 
\begin{equation} \label{AnGaussInt2}
\int\limits_{\mathbb{R}^N}
\int\limits_{-\infty}^{\mu_N - m_N} \cdots \int\limits_{-\infty}^{\mu_1 - m_1}\mathcal{N}(\mathbf{z}|\0, \mathrm{V} + \Sigma) \mathcal{N}(\mathbf{w}|c, C)\mathrm{d} \mathbf{z} \mathrm{d} \mathbf{w}
\end{equation}
where $c = - C \mathrm{V}^{-1} \mathbf{z}  $ and $C = (\mathrm{V}^{-1} + \Sigma^{-1})^{-1} $.
Interchange the order of integration with Fubini-Tonelli theorem \citep{folland:2013} and integrate w.r.t $\mathbf{w}$ to get that, 
\begin{equation} \label{AnGaussInt3}
\int\limits_{-\infty}^{\mu_N - m_N}\cdots\int\limits_{-\infty}^{\mu_1 - m_1} \mathcal{N}(\mathbf{z}|\0, \mathrm{V} + \Sigma)\mathrm{d}\mathbf{z} = F_N(\bmu_N \vert \mathbf{m}_N, \mathrm{V} + \Sigma)
\end{equation}
which completes the proof.
\end{proof}
The above closed-form integrals extend many equalities in the table of \cite{owen:1980}.

\bibliographystyle{ba}
\bibliography{refs}
\appendix

\newpage

\section{Extra results}

\begin{table}[!h]
\small
\begin{tabular}{l|cc}
\hline
Standard models  & LOO Cross Validation  & 5-fold Cross validation \\ \hline
\multirow{1}{6.1cm}{\textbf{1) Univariate GP $h(\x)$, univ. $\epsilon(\s)$} } &  \textbf{-2.230 (0.082)} $\phantom{-}$	& \textbf{-2.465 (0.094)} $\phantom{-}$  \\
$\phantom{-}$ diatom.algae  &  -3.942 (0.225)	& -4.002 (0.225) \\
$\phantom{-}$ filame.algae  &  -4.568 (0.141)	& -4.692 (0.146) \\
$\phantom{-}$ macro-veg     &  -1.909 (0.234)	& -2.092 (0.240) \\
$\phantom{-}$ threespine-sb &  -1.477 (0.157)	& -1.624 (0.154) \\
$\phantom{-}$ ninespine-sb  &  -1.334 (0.151)	& -1.464 (0.149) \\
$\phantom{-}$ white-fish    &  -2.840 (0.200)	& -3.092 (0.216) \\
$\phantom{-}$ vendance      &  -1.634 (0.210)	& -2.196 (0.314) \\
\multirow{1}{6.1cm}{\textbf{2) Univariate GP $h(\x)$, multiv. $\epsilon(\s)$} } &  \textbf{-2.081 (0.080)} $\phantom{-}$ & \textbf{-2.480 (0.100)} $\phantom{-}$ \\ 
$\phantom{-}$ diatom.algae  &  -3.578 (0.232)	& -4.010 (0.217) \\
$\phantom{-}$ filame.algae  &  -4.104 (0.159)	& -4.680 (0.135) \\
$\phantom{-}$ macro-veg     &  -1.733 (0.223)	& -2.148 (0.252) \\
$\phantom{-}$ threespine-sb &  -1.385 (0.149)	& -1.637 (0.156) \\
$\phantom{-}$ ninespine-sb  &  -1.232 (0.139)	& -1.489 (0.151) \\
$\phantom{-}$ white-fish    &  -2.782 (0.203)	& -3.092 (0.213) \\
$\phantom{-}$ vendance      &  -1.537 (0.204)	& -2.215 (0.316) \\
\multirow{1}{6.1cm}{\textbf{3) Multiv. GP $h(\x)$, multiv. $\epsilon(\s)$} }   &  \textbf{-1.669 (0.080)} $\phantom{-}$ & \textbf{-2.316 (0.086)} $\phantom{-}$ \\
$\phantom{-}$ diatom.algae  &  -2.307 (0.112)	& -3.947 (0.217) \\
$\phantom{-}$ filame.algae  &  -1.244 (0.751)	& -4.623 (0.140) \\
$\phantom{-}$ macro-veg     &  -1.289 (0.150)	& -1.950 (0.248) \\
$\phantom{-}$ threespine-sb &  -1.210 (0.140)	& -1.509 (0.156) \\
$\phantom{-}$ ninespine-sb  &  -1.184 (0.147)	& -1.389 (0.149) \\
$\phantom{-}$ white-fish    &  -3.399 (0.393)	& -2.964 (0.205) \\
$\phantom{-}$ vendance      &  -2.041 (0.451)	& -1.837 (0.247) \\ \hline
\multirow{1}{5.65cm}{\textbf{4) Multiv. GP $h(\x)$ only} } &  \textbf{ -2.228(0.089)} $\phantom{-}$ & \textbf{ -2.590(0.012)} $\phantom{-}$ \\ 
$\phantom{-}$ diatom.algae  &  -4.033(0.235)	& -4.545(0.379) \\
$\phantom{-}$ filame.algae  &  -4.697(0.156) 	& -4.865(0.187) \\
$\phantom{-}$ macro-veg     &  -1.754(0.226) 	& -2.514(0.379) \\
$\phantom{-}$ threespine-sb &  -1.418(0.173)	& -1.494(0.172)\\
$\phantom{-}$ ninespine-sb  &  -1.355(0.183) 	& -1.393(0.168) \\
$\phantom{-}$ white-fish    &  -2.866(0.212) 	& -3.192(0.271) \\
$\phantom{-}$ vendance      &  -1.608(0.222) 	& -2.458(0.400)\\
\multirow{1}{5.65cm}{\textbf{5) Multiv. $\epsilon(\s)$ only} }  &  \textbf{ -2.351(0.087)} $\phantom{-}$ & \textbf{-2.500(0.086)} $\phantom{-}$ \\
$\phantom{-}$ diatom.algae  &  -4.129(0.200) 	&  -4.231(0.182) \\
$\phantom{-}$ filame.algae  &  -4.607(0.119) 	&  -4.729(0.109) \\
$\phantom{-}$ macro-veg     &  -2.255(0.254) 	&  -2.800(0.210) \\
$\phantom{-}$ threespine-sb &  -1.471(0.174) 	&  -1.567(0.168) \\
$\phantom{-}$ ninespine-sb  &  -1.375(0.178) 	&  -1.473(0.152) \\
$\phantom{-}$ white-fish    &  -3.147(0.181) 	&  -3.226(0.191) \\
$\phantom{-}$ vendance      &  -1.675(0.235) 	&  -1.864(0.260) \\ \hline
\end{tabular} 
\caption{\label{tab:loocv} Model comparison with leave-one-out (LOO) and 5-fold cross validation using the mean point wise log marginal predictive density statistics (and its standard error) for models 1-5 (see Section~5.3 in the main text). The bolded numbers show the overall performance over all species and the indented text shows the species specific predictions.}
\end{table}

\begin{table}[!h]
\begin{tabular}{l|cc}
\hline
Standard models  & LOO Cross Validation  & 5-fold Cross validation \\ \hline
\multirow{1}{6.1cm}{\textbf{6) Univ. quadr. $h(\x)$, univ. $\epsilon(\s)$} } &  \textbf{-2.262 (0.083)} $\phantom{-}$	& \textbf{-2.517 (0.095)} $\phantom{-}$  \\
$\phantom{-}$ diatom.algae  &   -3.941(0.228)	&  -4.035(0.237) \\
$\phantom{-}$ filame.algae  &   -4.608(0.138)	&  -4.741(0.150) \\
$\phantom{-}$ macro-veg     &   -1.955(0.239)	&  -2.233(0.300) \\
$\phantom{-}$ threespine-sb &   -1.490(0.158)	&  -1.598(0.156) \\
$\phantom{-}$ ninespine-sb  &   -1.337(0.151)	&  -1.501(0.140) \\
$\phantom{-}$ white-fish    &   -2.885(0.204)	&  -3.137(0.215) \\
$\phantom{-}$ vendance      &   -1.705(0.215)	&  -2.274(0.315) \\
\multirow{1}{6.1cm}{\textbf{7) Univ. quadr. $h(\x)$, multiv. $\epsilon(\s)$} } &  \textbf{-2.117 (0.084)} $\phantom{-}$ & \textbf{-2.484 (0.109)} $\phantom{-}$ \\ 
$\phantom{-}$ diatom.algae  &   -3.568(0.243)	& -3.994(0.225) \\
$\phantom{-}$ filame.algae  &   -4.104(0.170)	& -4.717(0.137) \\
$\phantom{-}$ macro-veg     &   -1.763(0.229)	& -2.269(0.294) \\
$\phantom{-}$ threespine-sb &   -1.360(0.165)	& -1.517(0.171) \\
$\phantom{-}$ ninespine-sb  &   -1.233(0.143)	& -1.484(0.142) \\
$\phantom{-}$ white-fish    &   -2.853(0.210)	& -3.098(0.251) \\
$\phantom{-}$ vendance      &   -1.668(0.226)	& -2.208(0.340) \\
\multirow{1}{6.1cm}{\textbf{8) Multiv. quadr. $h(\x)$, multiv. $\epsilon(\s)$} }   &  \textbf{-2.105 (0.084)} $\phantom{-}$ & \textbf{-2.416 (0.099)} $\phantom{-}$ \\
$\phantom{-}$ diatom.algae  &   -3.561(0.242)	&  -4.009(0.242) \\
$\phantom{-}$ filame.algae  &   -4.090(0.177)	&  -4.745(0.170) \\
$\phantom{-}$ macro-veg     &   -1.742(0.229)	&  -2.197(0.280) \\
$\phantom{-}$ threespine-sb &   -1.346(0.165)	&  -1.476(0.173) \\
$\phantom{-}$ ninespine-sb  &   -1.217(0.149)	&  -1.441(0.146) \\
$\phantom{-}$ white-fish    &   -2.845(0.201)	&  -3.045(0.241) \\
$\phantom{-}$ vendance      &   -1.663(0.226)	&  -2.080(0.311) \\
\multirow{1}{5.65cm}{\textbf{9) Multiv. quadr. $h(\x)$ only} } &  \textbf{ -10.246(1.297)} $\phantom{-}$	& \textbf{ -9.305(1.008)} $\phantom{-}$  \\
$\phantom{-}$ diatom.algae  &  -48.063(10.943)	& -32.673(6.132) \\
$\phantom{-}$ filame.algae  &  -52.309(8.539)	& -42.283(6.498) \\
$\phantom{-}$ macro-veg     &  -7.139(1.439)	& -19.895(6.421) \\
$\phantom{-}$ threespine-sb &  -1.441(0.178)	& -1.494(0.177)  \\
$\phantom{-}$ ninespine-sb  &  -1.375(0.192)	& -1.400(0.158)  \\
$\phantom{-}$ white-fish    &  -2.899(0.215)	& -3.209(0.275)  \\
$\phantom{-}$ vendance      &  -1.677(0.227)	& -2.311(0.364)  \\
\end{tabular} 
\caption{\label{tab:loocv} Model comparison with leave-one-out (LOO) and 5-fold cross validation using the mean point wise log marginal predictive density statistics (and its standard error) for models 6-9 (see Section~5.3 in the main text). The bolded numbers show the overall performance over all species and the indented text shows the species specific predictions.}
\end{table}
\begin{table}[!h]
\small
\begin{tabular}{l|rr}
\hline
Pairwise model comparison & LOO Cross Validation  & 5-fold Cross validation \\ \hline
\multirow{1}{5.65cm}{\textbf{Model~8 - Model~6} } &  \textbf{ 0.156(0.019)} $\phantom{-}$	& \textbf{ 0.101(0.030) } $\phantom{-}$  \\
$\phantom{-}$ diatom.algae  & 0.379(0.106) 	&  0.025(0.020)  \\
$\phantom{-}$ filame.algae  & 0.537(0.097) 	& -0.003(0.037)  \\
$\phantom{-}$ macro-veg     & 0.214(0.214) 	&  0.135(0.065)  \\
$\phantom{-}$ threespine-sb & 0.144(0.032) 	&  0.012(0.023) \\
$\phantom{-}$ ninespine-sb  & 0.120(0.026) 	&  0.059(0.017) \\
$\phantom{-}$ white-fish    & 0.036(0.039)	&  0.091(0.045) \\
$\phantom{-}$ vendance      & 0.043(0.032) 	&  0.194(0.031) \\
\multirow{1}{5.65cm}{\textbf{Model~8 - Model~7}} &  \textbf{ 0.012(0.004)} $\phantom{-}$ & \textbf{ 0.052(0.010)} $\phantom{-}$ \\ 
$\phantom{-}$ diatom.algae  & 0.006(0.016) 	& -0.015(0.028)  \\
$\phantom{-}$ filame.algae  & 0.014(0.021)	& -0.027(0.046) \\
$\phantom{-}$ macro-veg     & 0.021(0.012)	& 0.072(0.028) \\
$\phantom{-}$ threespine-sb & 0.015(0.008)	& 0.041(0.015) \\
$\phantom{-}$ ninespine-sb  & 0.016(0.011) 	& 0.043(0.017) \\
$\phantom{-}$ white-fish    & 0.008(0.011) 	& 0.052(0.019) \\
$\phantom{-}$ vendance      & 0.006(0.006)	& 0.128(0.033) \\ \hline
\end{tabular} 
\caption{\label{tab:loocv} Pairwise comparison of leave-one-out (LOO) and 5-fold cross validation point wise log predictive densities of multivariate parametric (quadratic environmental responses) model (model~8) and parametric models without any interspecific correlations (model~6) or with interspecific correlations only in spatial random effect (model~7). We report the mean (and its standard error) of the differences in point wise log predictive densities at test locations. The bolded numbers show the overall performance and normal text shows the species specific predictions.}
\end{table}
\begin{table}[!h]
\small
\begin{tabular}{l|rr}
\hline
Pairwise model comparison & LOO Cross Validation  & 5-fold Cross validation \\ \hline
\multirow{1}{5cm}{\textbf{Model~3 - model~1}} & \textbf{0.561(0.126)} $\phantom{-}$	& \textbf{ 0.150(0.123) } $\phantom{-}$  \\
$\phantom{-}$ diatom.algae  &  1.637(0.151) 	& 0.054(0.029)  \\
$\phantom{-}$ filame.algae  &  3.974(0.495)  	& 0.070(0.027)  \\
$\phantom{-}$ macro-veg     &  0.620(0.129) 	& 0.141(0.029)  \\
$\phantom{-}$ threespine-sb &  0.267(0.097) 	& 0.115(0.017)  \\
$\phantom{-}$ ninespine-sb  &  0.151(0.097) 	& 0.075(0.019) \\
$\phantom{-}$ white-fish    & -0.300(0.127) 	& 0.129(0.026)  \\
$\phantom{-}$ vendance      &  0.051(0.073) 	& 0.359(0.077)  \\
\multirow{1}{5.0cm}{\textbf{Model~3 - model~2}} & \textbf{ 0.413(0.123)} $\phantom{-}$ & \textbf{0.164(0.017)} $\phantom{-}$ \\ 
$\phantom{-}$ diatom.algae  &  1.271(0.136) & 0.065(0.029) \\
$\phantom{-}$ filame.algae  &  3.496(0.504) & 0.057(0.020) \\
$\phantom{-}$ macro-veg     &  0.444(0.011) & 0.198(0.031) \\
$\phantom{-}$ threespine-sb &  0.175(0.087) & 0.128(0.198) \\
$\phantom{-}$ ninespine-sb  &  0.049(0.085) & 0.100(0.021) \\
$\phantom{-}$ white-fish    & -0.357(0.011) & 0.128(0.026) \\
$\phantom{-}$ vendance      & -0.041(0.055) & 0.377(0.078) \\ \hline
\end{tabular} 
\caption{\label{tab:loocv} Pairwise comparison of leave-one-out (LOO) and 5-fold cross validation point wise log predictive densities of multivariate GP model (model~3) and GP models without any interspecific correlations (model~1) or with interspecific correlations only in spatial random effect (model~2). We report the mean (and its standard error) of the differences in point wise log predictive densities at test locations. The bolded numbers show the overall performance and normal text shows the species specific predictions.}
\end{table}

\begin{landscape}
\begin{table}
\caption{Hyperparameters estimates from covariates covariance functions}
\begin{tabular}{lrrrrrrr}\toprule
                          &  Diatom-algae & Filam-algae & Macro-veg & Threespine-sb & Ninespine-sb & White-fish & Vendance \\ 
                          & \scriptsize{SDM $\mid$ JSDM} & \scriptsize{SDM $\mid$ JSDM} & \scriptsize{SDM $\mid$ JSDM} & \scriptsize{SDM $\mid$ JSDM} & \scriptsize{SDM $\mid$ JSDM} & \scriptsize{SDM $\mid$ JSDM} & \scriptsize{SDM $\mid$ JSDM} \\[0.1cm]                
\midrule

\multirow{2}{*}{Opennes} & $\sigma^2$ \hspace{0.1cm} $1.92$ $\mid$ $2.19$ & $1.96$ $\mid$ $1.87$ & $1.63$ $\mid$ $1.62$ & $1.70$ $\mid$ $1.64$ & $2.18$ $\mid$ $1.78$ & $1.57$ $\mid$ $1.30$ & $2.61$ $\mid$ $2.47$ \\
                         & $\ell^{\phantom{2}}$ \hspace{0.1cm} $1.56$ $\mid$ $1.59$ & $1.29$ $\mid$ $1.31$ & $1.51$ $\mid$ $1.26$ & $1.90$ $\mid$ $2.23$ & $2.12$ $\mid$ $1.86$ & $1.98$ $\mid$ $1.61$ & $0.64$ $\mid$ $0.92$ \\
\midrule                         

\multirow{2}{*}{\parbox{1.9cm}{Distance to \\ deep}} & $\sigma^2$ \hspace{0.1cm} $2.24$ $\mid$ $1.84$ & $1.55$ $\mid$ $1.48$ & $2.87$ $\mid$ $2.64$ & $2.54$ $\mid$ $2.23$ & $1.98$ $\mid$ $1.43$ & $1.66$ $\mid$ $1.32$ & $1.79$ $\mid$ $1.37$ \\
                          & $\ell^{\phantom{2}}$  \hspace{0.1cm} $1.09$ $\mid$ $1.18$ & $1.16$ $\mid$ $1.06$ & $1.39$ $\mid$ $1.87$ & $1.04$ $\mid$ $1.39$ & $1.48$ $\mid$ $1.80$ & $2.02$ $\mid$ $1.33$ & $1.32$ $\mid$ $1.09$ \\ \midrule
\multirow{2}{*}{\parbox{2.2cm}{Sandy bottom index}} & $\sigma^2$ \hspace{0.1cm} $1.89$ $\mid$ $1.72$ & $1.51$ $\mid$ $1.63$ & $2.14$ $\mid$ $1.85$ & $2.11$ $\mid$ $1.49$ & $1.89$ $\mid$ $1.79$ & $3.37$ $\mid$ $2.26$ & $2.59$ $\mid$ $2.37$ \\
                          & $\ell^{\phantom{2}}$  \hspace{0.1cm} $1.49$ $\mid$ $1.43$ & $1.48$ $\mid$ $1.24$ & $1.30$ $\mid$ $0.84$ & $0.71$ $\mid$ $0.46$ & $1.93$ $\mid$ $1.60$ & $1.15$ $\mid$ $0.95$ & $0.91$ $\mid$ $1.06$ \\  \midrule
\multirow{2}{*}{\parbox{1.8cm}{Ice breakup week}} & $\sigma^2$ \hspace{0.1cm} $2.36$ $\mid$ $2.37$ & $2.16$ $\mid$ $1.74$ & $3.26$ $\mid$ $2.54$ & $1.70$ $\mid$ $1.63$ & $1.99$ $\mid$ $1.86$ & $1.64$ $\mid$ $1.09$ & $6.69$ $\mid$ $7.52$ \\
                          & $\ell^{\phantom{2}}$  \hspace{0.1cm} $1.11$ $\mid$ $1.18$ & $0.75$ $\mid$ $1.06$ & $0.62$ $\mid$ $1.87$ & $1.32$ $\mid$ $1.39$ & $2.22$ $\mid$ $1.80$ & $1.87$ $\mid$ $1.33$ & $0.70$ $\mid$ $1.09$ \\  \midrule

\multirow{2}{*}{Chlorophyl-a} & $\sigma^2$ \hspace{0.1cm} $1.62$ $\mid$ $1.53$ & $1.55$ $\mid$ $1.64$ & $2.93$ $\mid$ $2.68$ & $1.91$ $\mid$ $1.46$ & $1.72$ $\mid$ $1.42$ & $1.95$ $\mid$ $1.75$ & $2.53$ $\mid$ $2.48$ \\
                          & $\ell^{\phantom{2}}$  \hspace{0.1cm} $1.11$ $\mid$ $1.46$ & $0.75$ $\mid$ $0.78$ & $0.62$ $\mid$ $0.49$ & $1.32$ $\mid$ $1.76$ & $2.22$ $\mid$ $1.71$ & $1.87$ $\mid$ $1.30$ & $0.70$ $\mid$ $0.66$ \\ \midrule

\multirow{2}{*}{\parbox{1.9cm}{Distance to \\ nearest river}} & $\sigma^2$ \hspace{0.1cm} $1.68$ $\mid$ $1.42$ & $1.56$ $\mid$ $1.30$ & $1.69$ $\mid$ $2.26$ & $1.96$ $\mid$ $1.71$ & $1.90$ $\mid$ $2.34$ & $1.67$ $\mid$ $2.67$ & $1.65$ $\mid$ $1.55$ \\
                          & $\ell^{\phantom{2}}$ \hspace{0.1cm} $1.59$ $\mid$ $1.76$ & $1.63$ $\mid$ $1.42$ & $1.10$ $\mid$ $0.80$ & $0.76$ $\mid$ $0.65$ & $1.23$ $\mid$ $0.73$ & $0.36$ $\mid$ $0.41$ & $1.45$ $\mid$ $1.63$ \\ \midrule

\multirow{2}{*}{Bottom type} & $\sigma^2$ \hspace{0.1cm} $0.81$ $\mid$ $1.01$ & $0.69$ $\mid$ $1.07$ & $0.69$ $\mid$ $0.81$ & $1.77$ $\mid$ $0.87$ & $1.08$ $\mid$ $1.00$ & $2.01$ $\mid$ $1.00$ & $0.75$ $\mid$ $1.00$ \\
                          & $\phantom{0.00}$ $\mid$ $\phantom{0.00}$ & $\phantom{0.00}$ $\mid$ $\phantom{0.00}$ & $\phantom{0.00}$ $\mid$ $\phantom{0.00}$ & $\phantom{0.00}$ $\mid$ $\phantom{0.00}$ & $\phantom{0.00}$ $\mid$ $\phantom{0.00}$ & $\phantom{0.00}$ $\mid$ $\phantom{0.00}$ & $\phantom{0.00}$ $\mid$ $\phantom{0.00}$ \\ \midrule

\multirow{2}{*}{Salinity} & $\sigma^2$ \hspace{0.1cm} $\phantom{0.00}$ $\mid$ $\phantom{0.00}$ & $\phantom{0.00}$ $\mid$ $\phantom{0.00}$ & $\phantom{0.00}$ $\mid$ $\phantom{0.00}$ & $\phantom{0.00}$ $\mid$ $\phantom{0.00}$ & $\phantom{0.00}$ $\mid$ $\phantom{0.00}$ & $\phantom{0.00}$ $\mid$ $\phantom{0.00}$ & $2.87$ $\mid$ $2.17$ \\
                          & $\ell^{\phantom{2}}$ \hspace{0.1cm} $\phantom{0.00}$ $\mid$ $\phantom{0.00}$ & $\phantom{0.00}$ $\mid$ $\phantom{0.00}$ & $\phantom{0.00}$ $\mid$ $\phantom{0.00}$ & $\phantom{0.00}$ $\mid$ $\phantom{0.00}$ & $\phantom{0.00}$ $\mid$ $\phantom{0.00}$ & $\phantom{0.00}$ $\mid$ $\phantom{0.00}$ & $1.24$ $\mid$ $1.37$ \\\midrule

\multirow{2}{*}{Spatial} & $\sigma^2$ \hspace{0.1cm} $10.9$ $\mid$ $9.35$ & $7.31$ $\mid$ $6.00$ & $4.62$ $\mid$ $4.67$ & $6.87$ $\mid$ $6.42$ & $5.66$ $\mid$ $5.74$ & $8.11$ $\mid$ $7.39$ & $9.32$ $\mid$ $6.85$ \\
                         & $\ell^{\phantom{2}}$ \hspace{0.1cm} $0.47$ $\mid$ $0.45$ & $0.74$ $\mid$ $1.06$ & $0.61$ $\mid$ $0.44$ & $0.44$ $\mid$ $0.43$ & $0.78$ $\mid$ $0.95$ & $0.15$ $\mid$ $0.22$ & $2.14$ $\mid$ $2.45$ \\ \bottomrule\end{tabular}
\end{table}
\end{landscape}


\begin{figure}[!htb]
\includegraphics[scale=0.75]{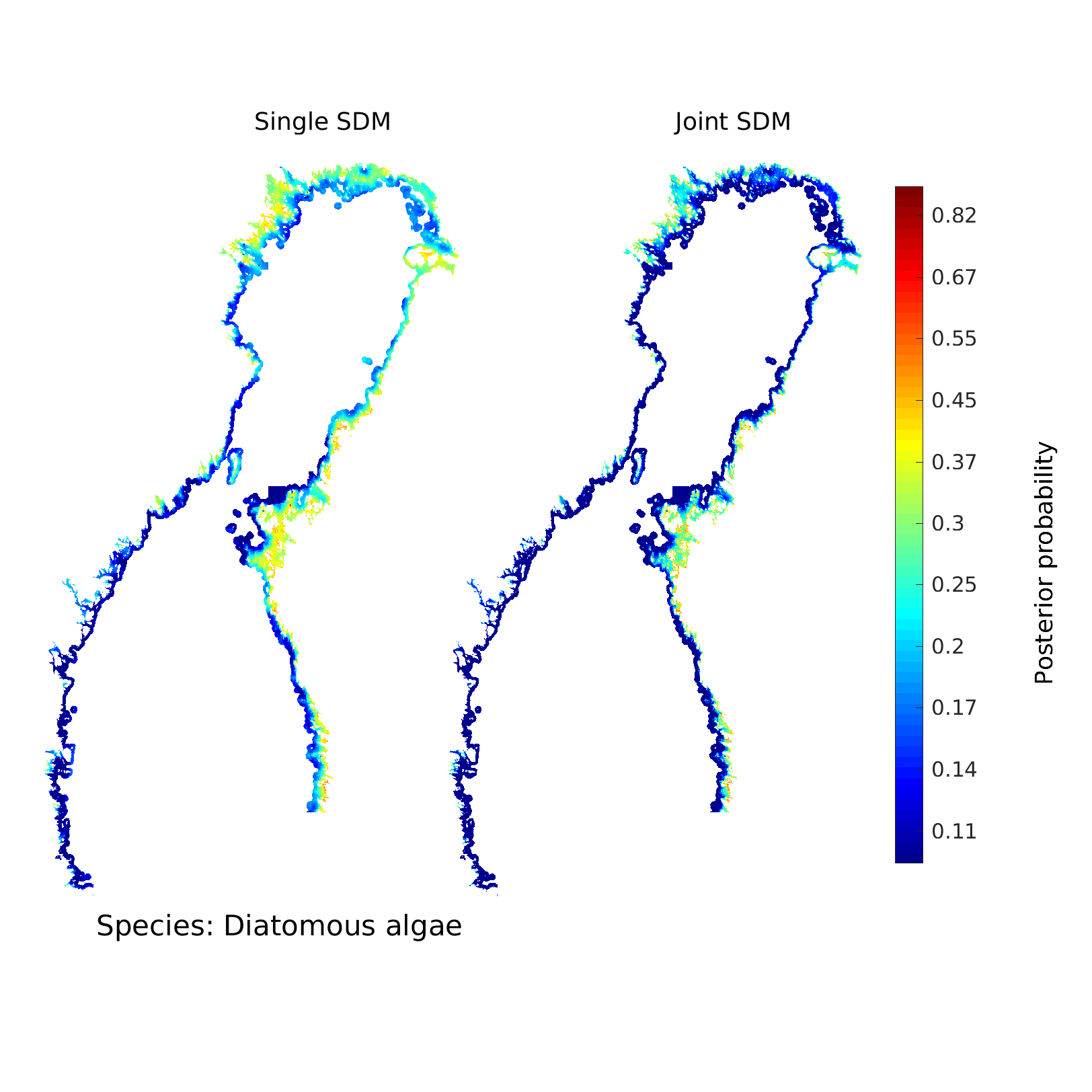}
\caption{Posterior predictive median of intensity and the expected count of individuals in replicate sampling for diatomo algae predicted by SSDM (model 1) and JSDM (model 3).}
\end{figure}

\begin{figure}[!htb]
\includegraphics[scale=0.75]{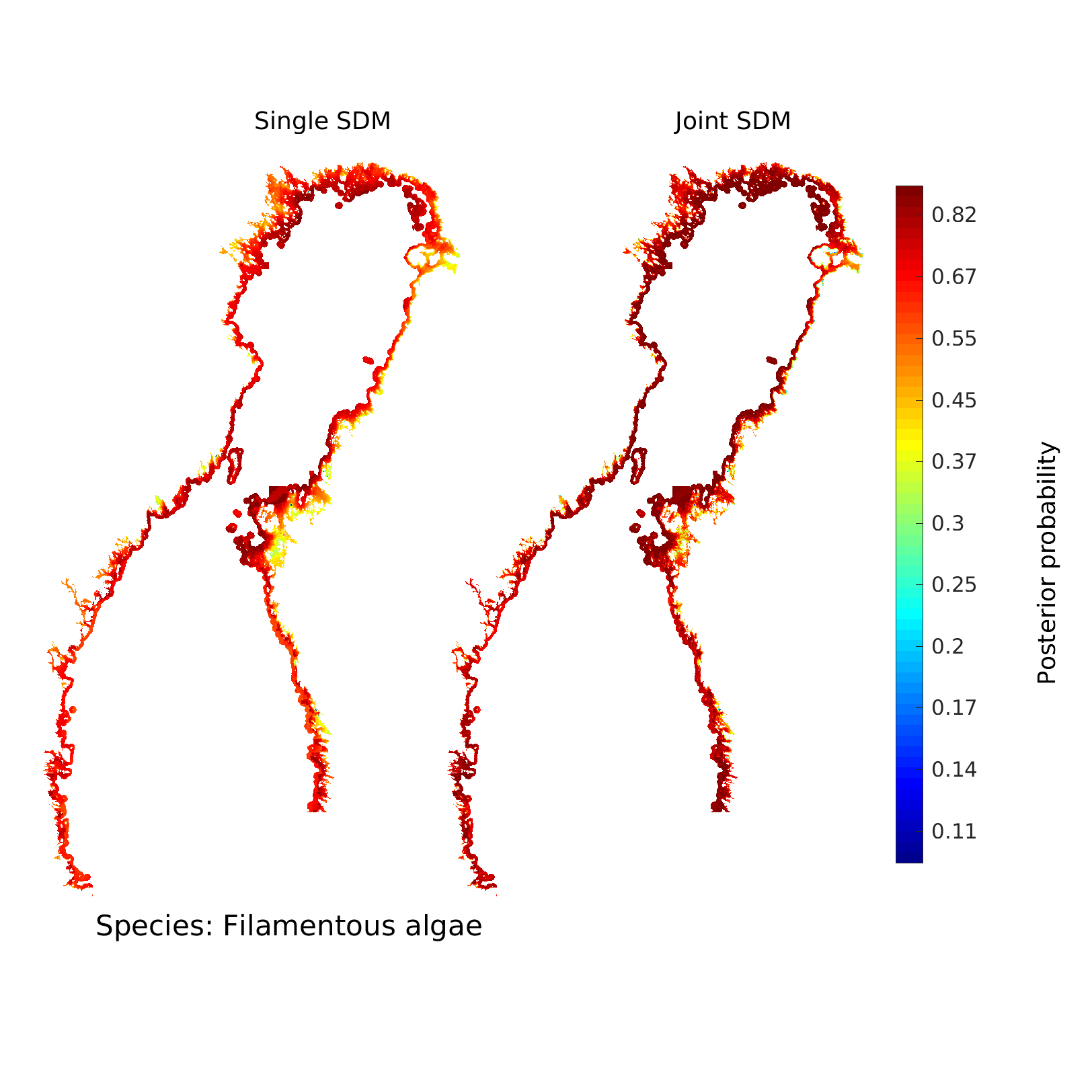}
\caption{Posterior predictive median of intensity and the expected count of individuals in replicate sampling for filamentous algae predicted by SSDM (model 1) and JSDM (model 3).}
\end{figure}

\begin{figure}[!htb]
\includegraphics[scale=0.75]{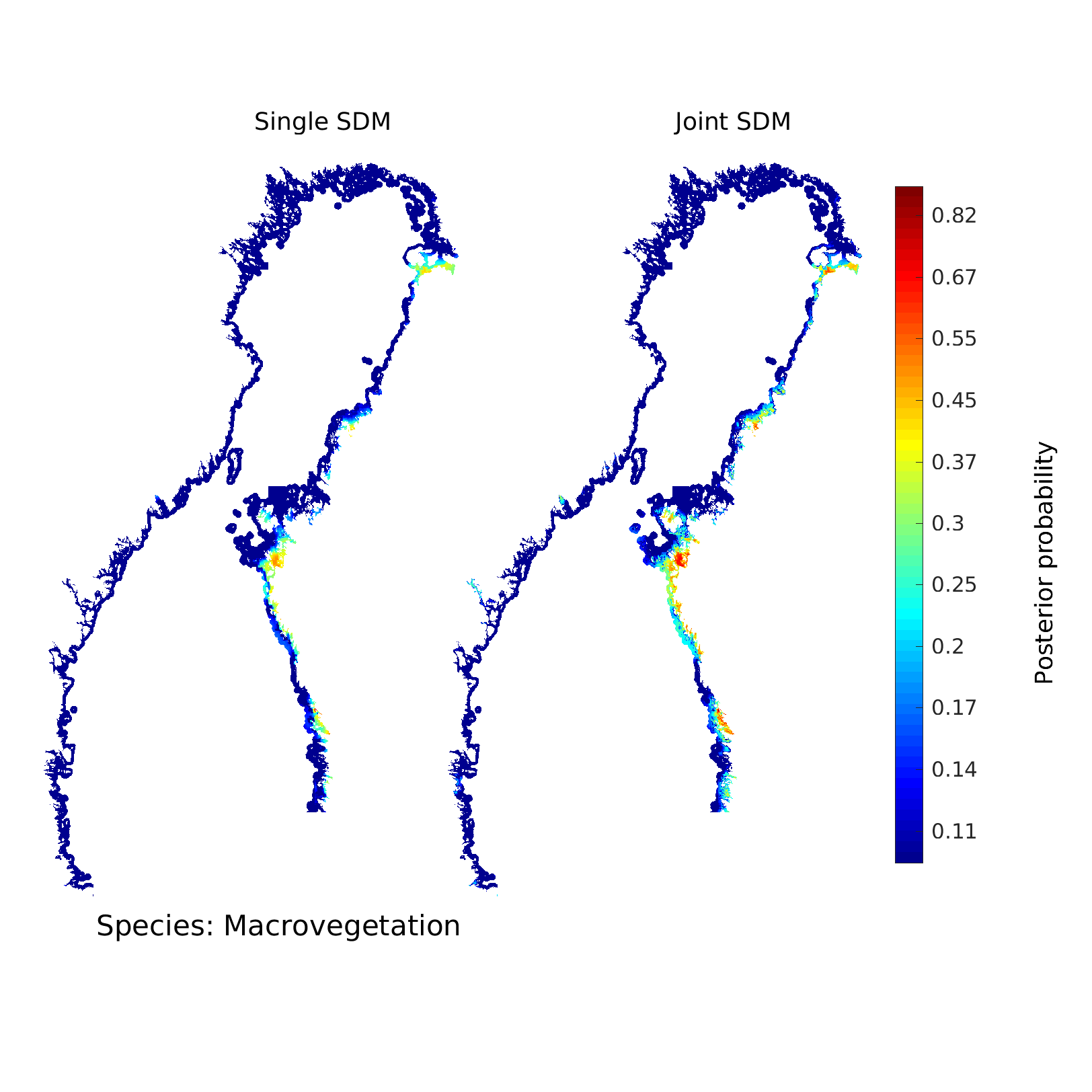}
\caption{Posterior predictive median of intensity and the expected count of individuals in replicate sampling for macrovegetation predicted by SSDM (model 1) and JSDM (model 3).}
\end{figure}

\begin{figure}[!htb]
\includegraphics[scale=0.9]{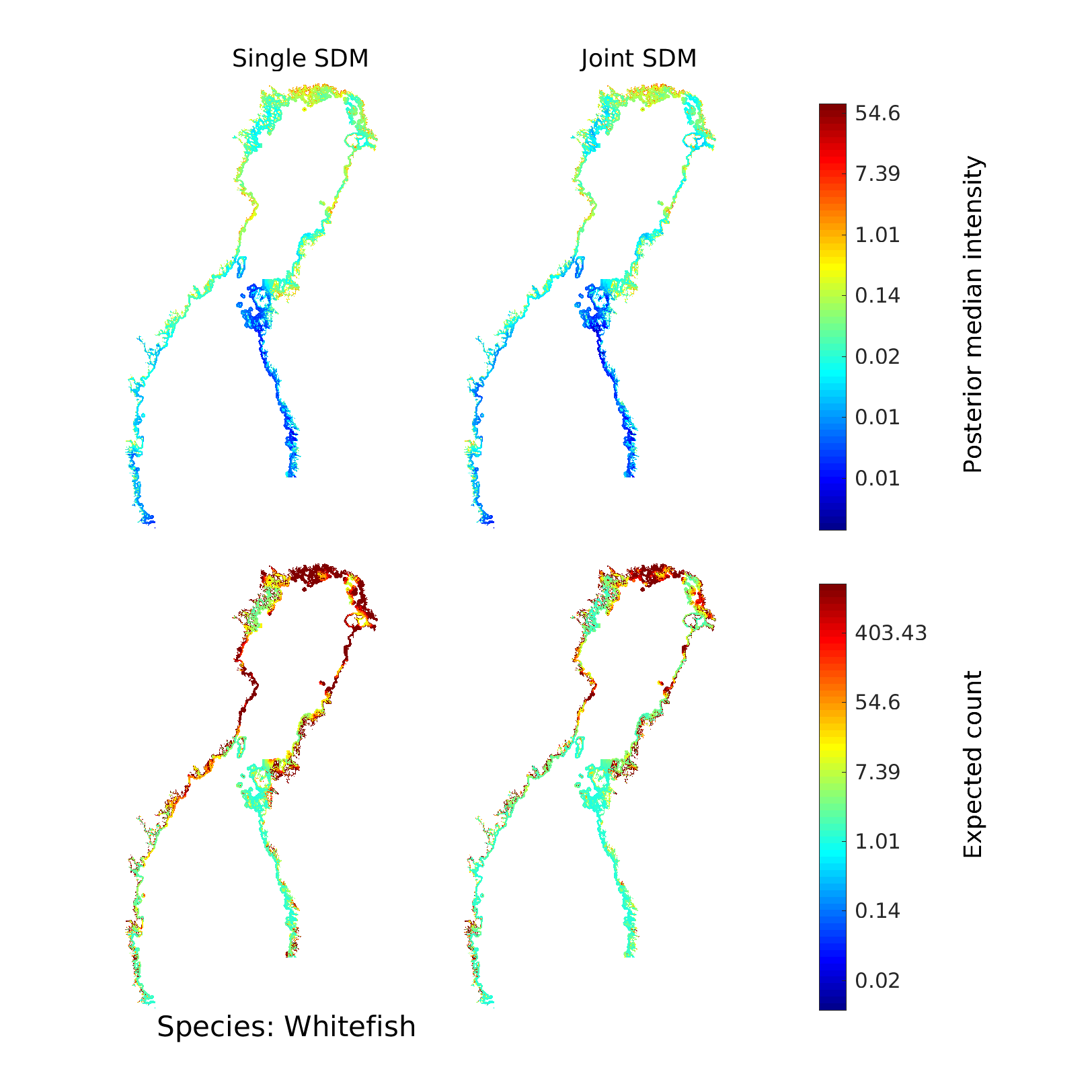}
\caption{Posterior predictive median of intensity and the expected count of individuals in replicate sampling for whitefish predicted by SSDM (model 1) and JSDM (model 3).}
\end{figure}

\begin{figure}[!htb]
\includegraphics[scale=0.9]{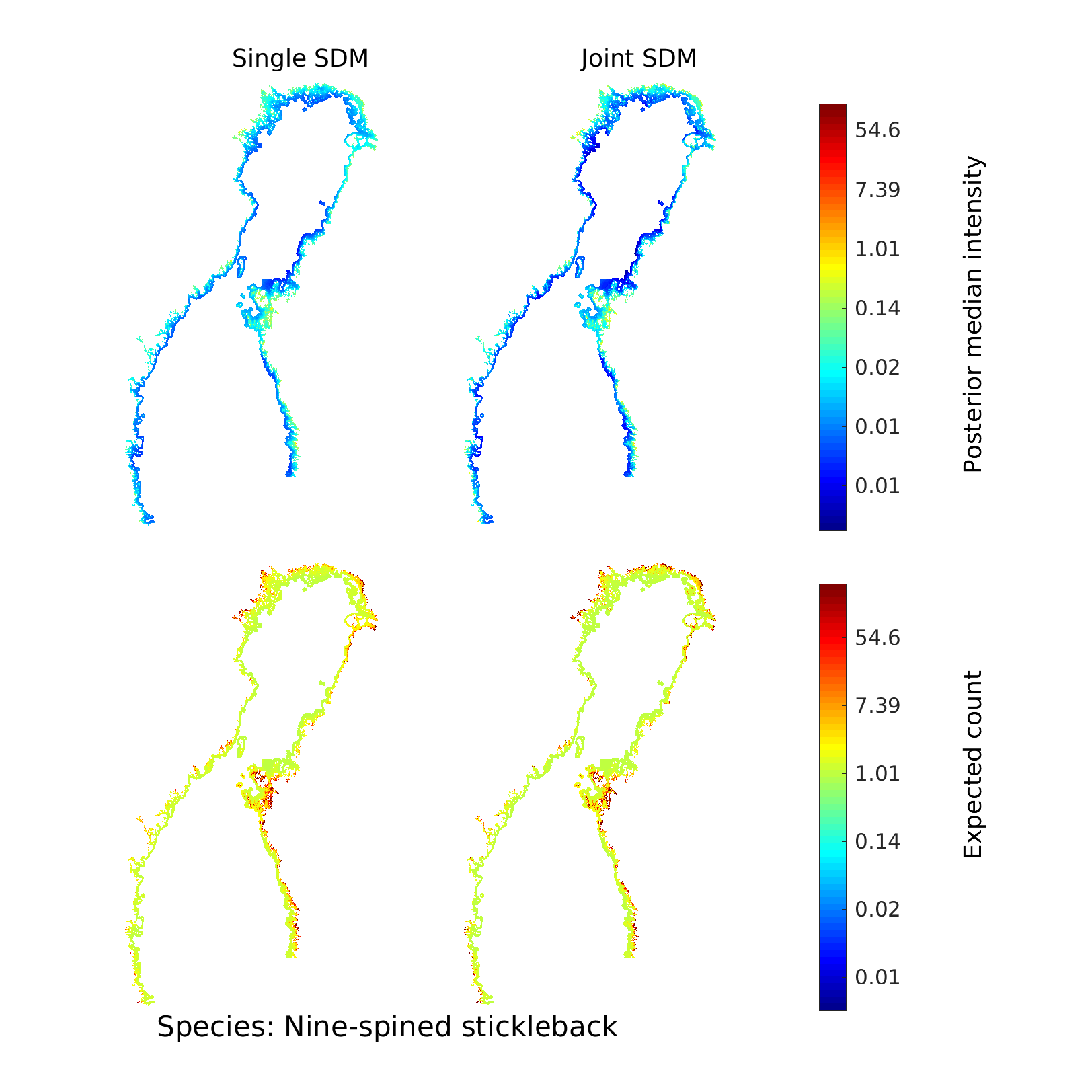}
\caption{Posterior predictive median of intensity and the expected count of individuals in replicate sampling for ninespine-stickleback predicted by SSDM (model 1) and JSDM (model 3).}
\end{figure}
\begin{figure}[!htb]
\includegraphics[scale=0.9]{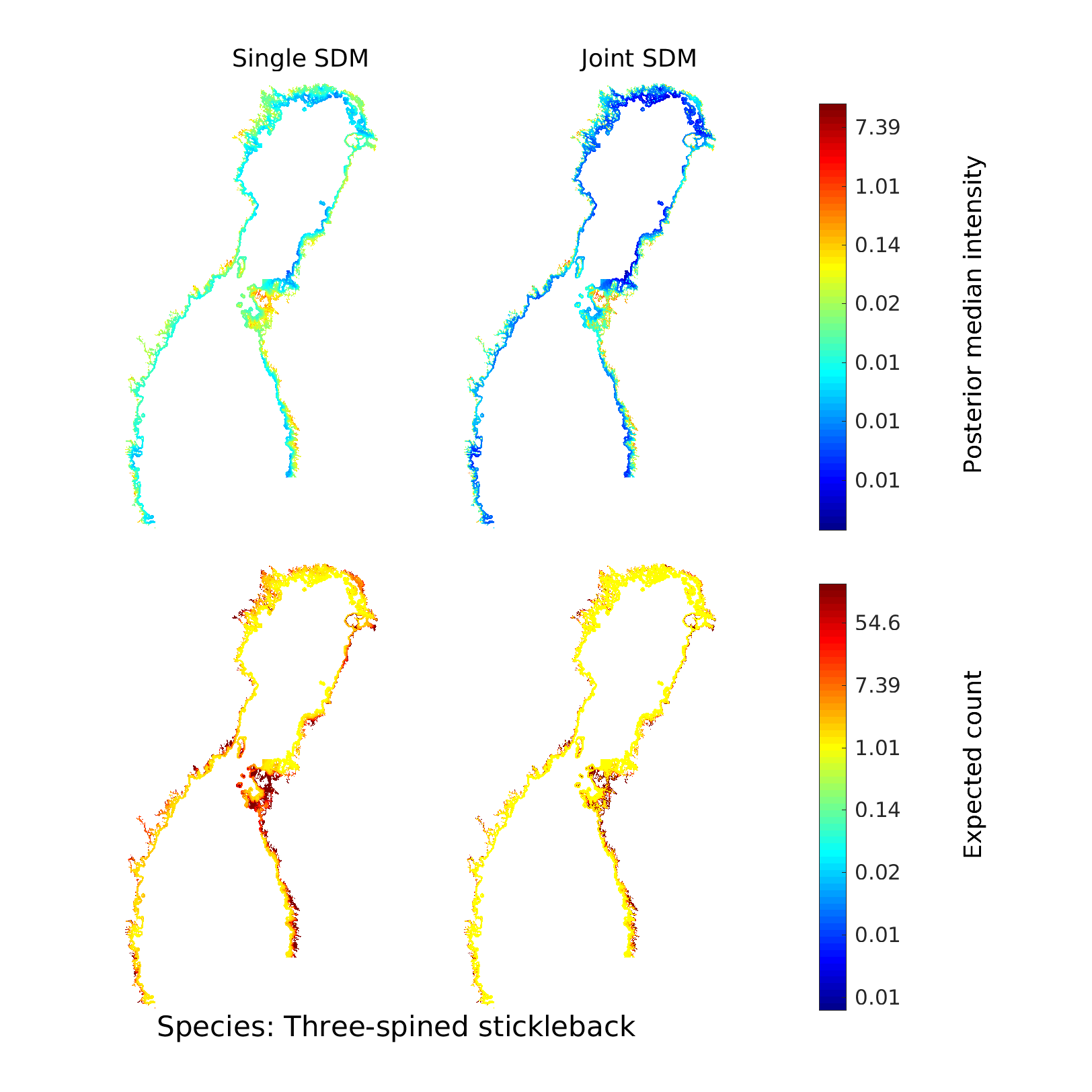}
\caption{Posterior predictive median of intensity and the expected count of individuals in replicate sampling for threespine-stickleback predicted by SSDM (model 1) and JSDM (model 3).}
\end{figure}


\begin{figure}[!htb]
\includegraphics[scale=0.75]{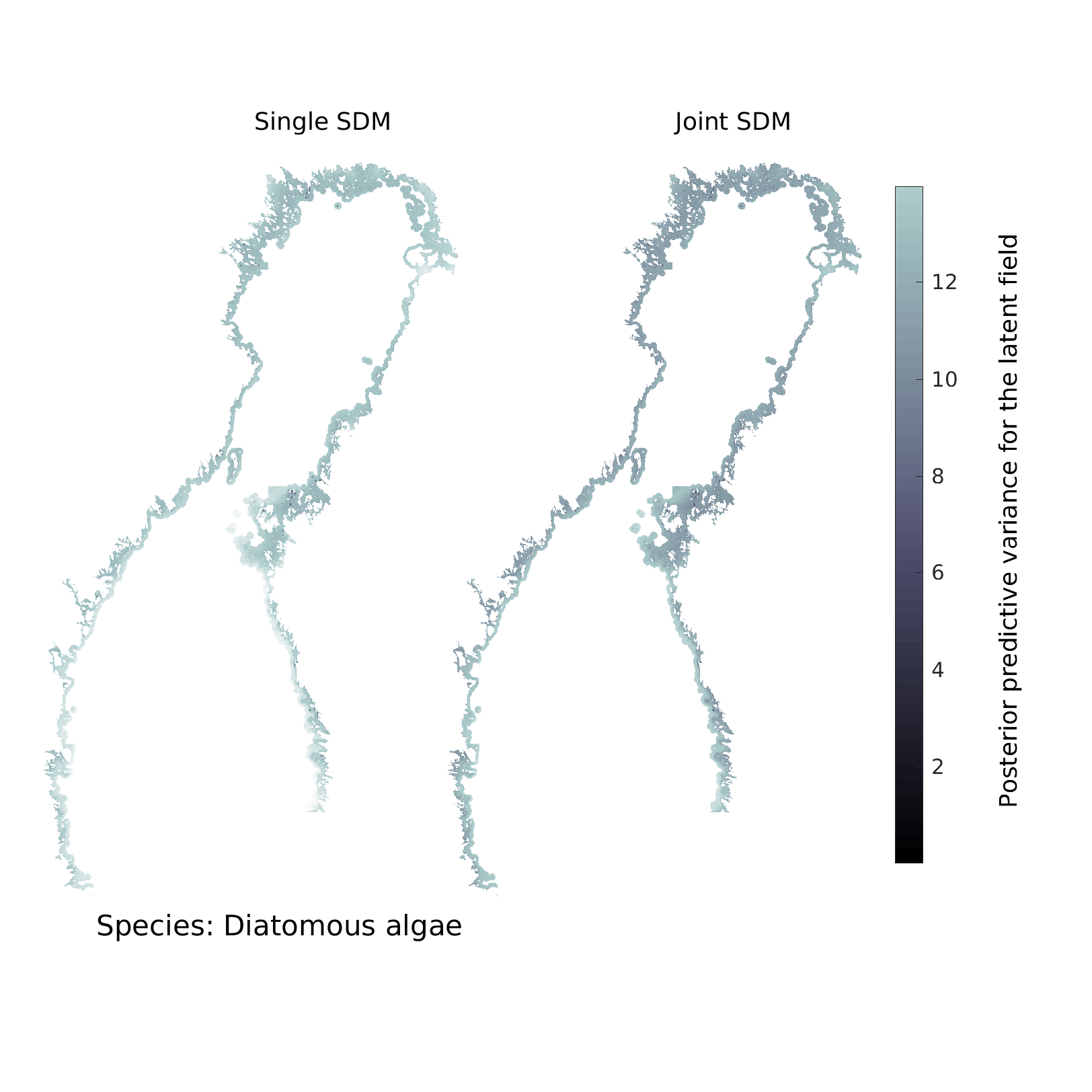}
\caption{Posterior predictive variance for the latent field $f_*(\x, \s)|\y, \hat{\eta}, \hat{\lambda}$ diatomo algae predicted by SSDM (model 1) and JSDM (model 3).}
\end{figure}

\begin{figure}[!htb]
\includegraphics[scale=0.75]{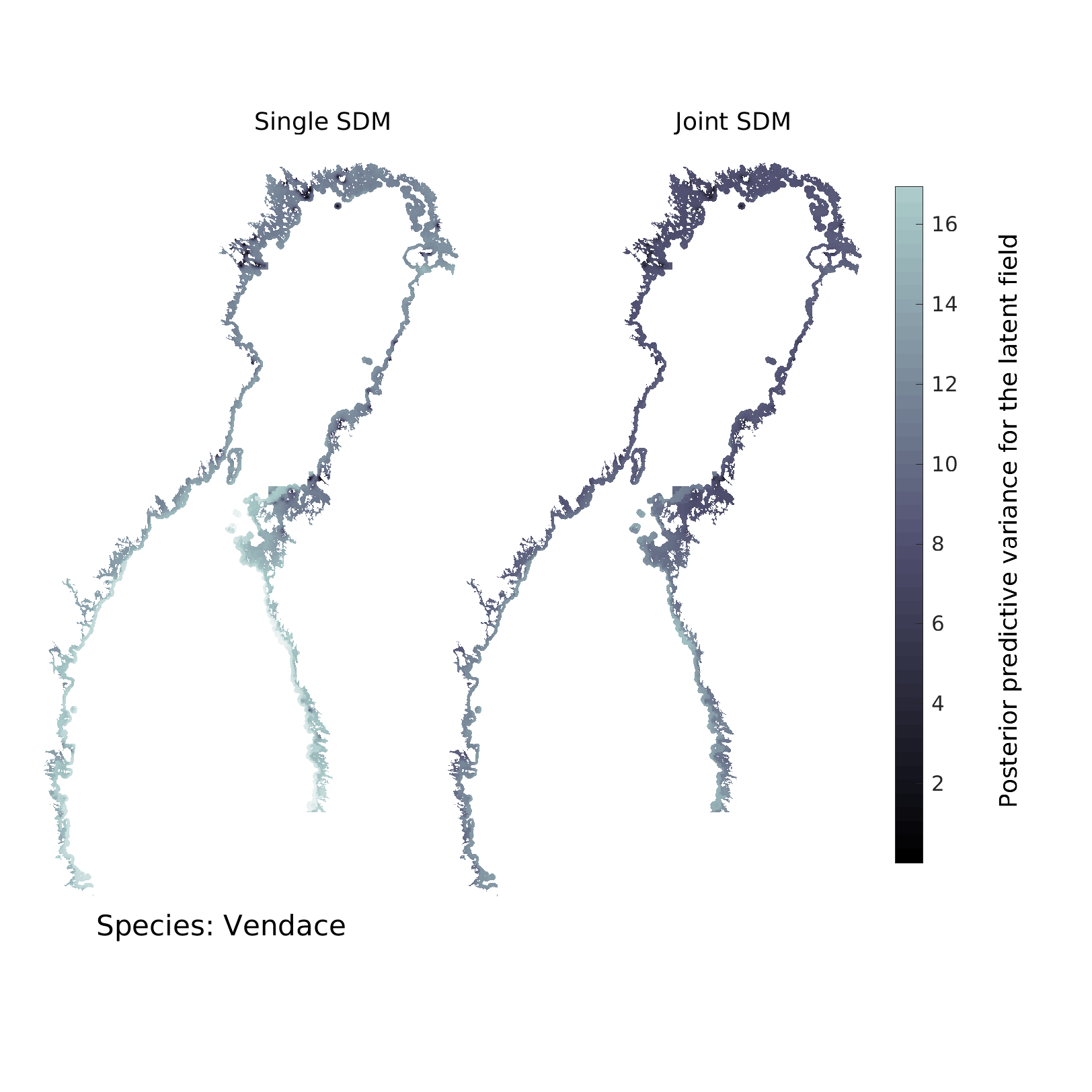}
\caption{Posterior predictive variance for the latent field $f_*(\x, \s)|\y, \hat{\eta}, \hat{\lambda}$ vendace predicted by SSDM (model 1) and JSDM (model 3).}
\end{figure}

\begin{figure}[!htb]
\includegraphics[scale=0.75]{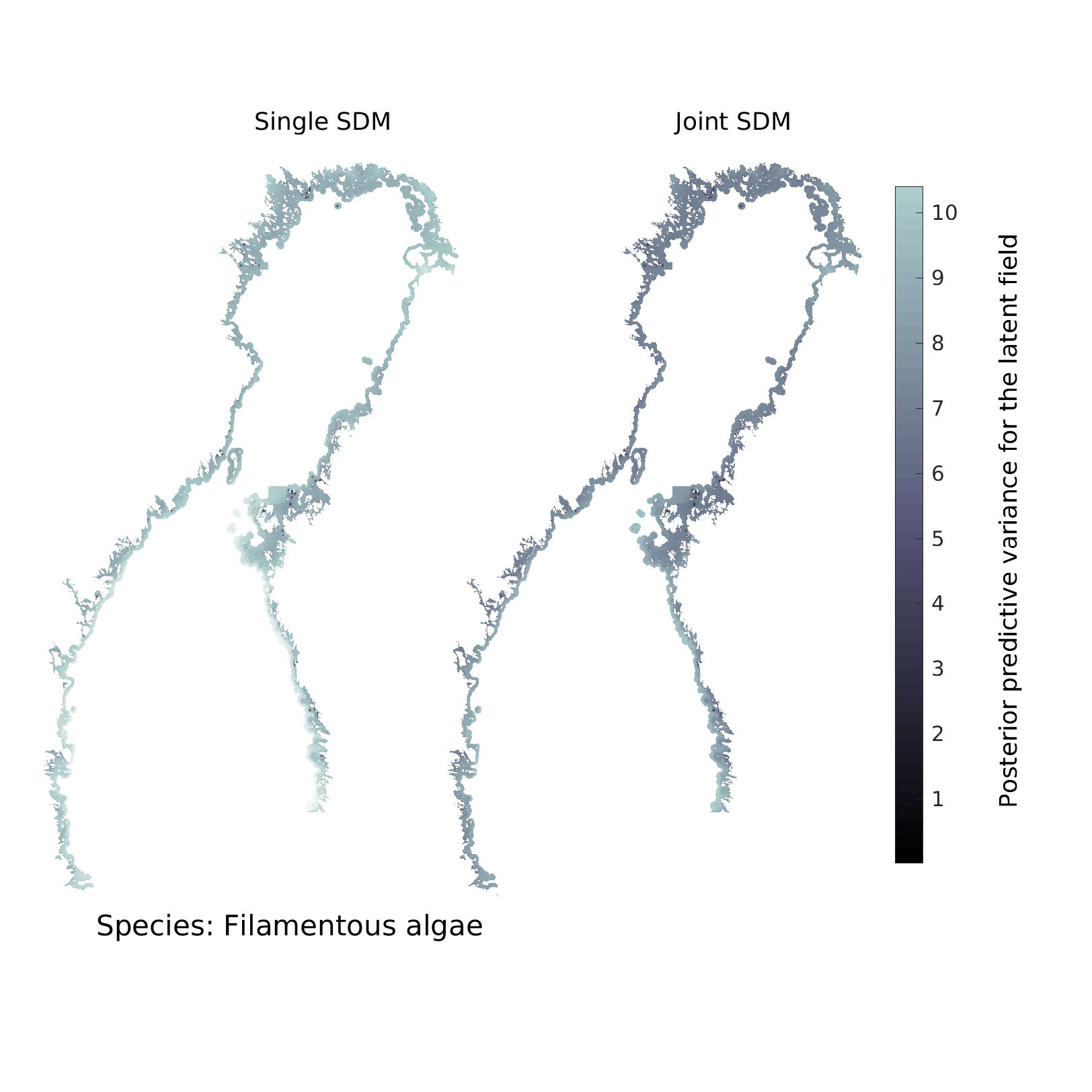}
\caption{Posterior predictive variance for the latent field $f_*(\x, \s)|\y, \hat{\eta}, \hat{\lambda}$ for filamentous algae predicted by SSDM (model 1) and JSDM (model 3).}
\end{figure}

\begin{figure}[!htb]
\includegraphics[scale=0.75]{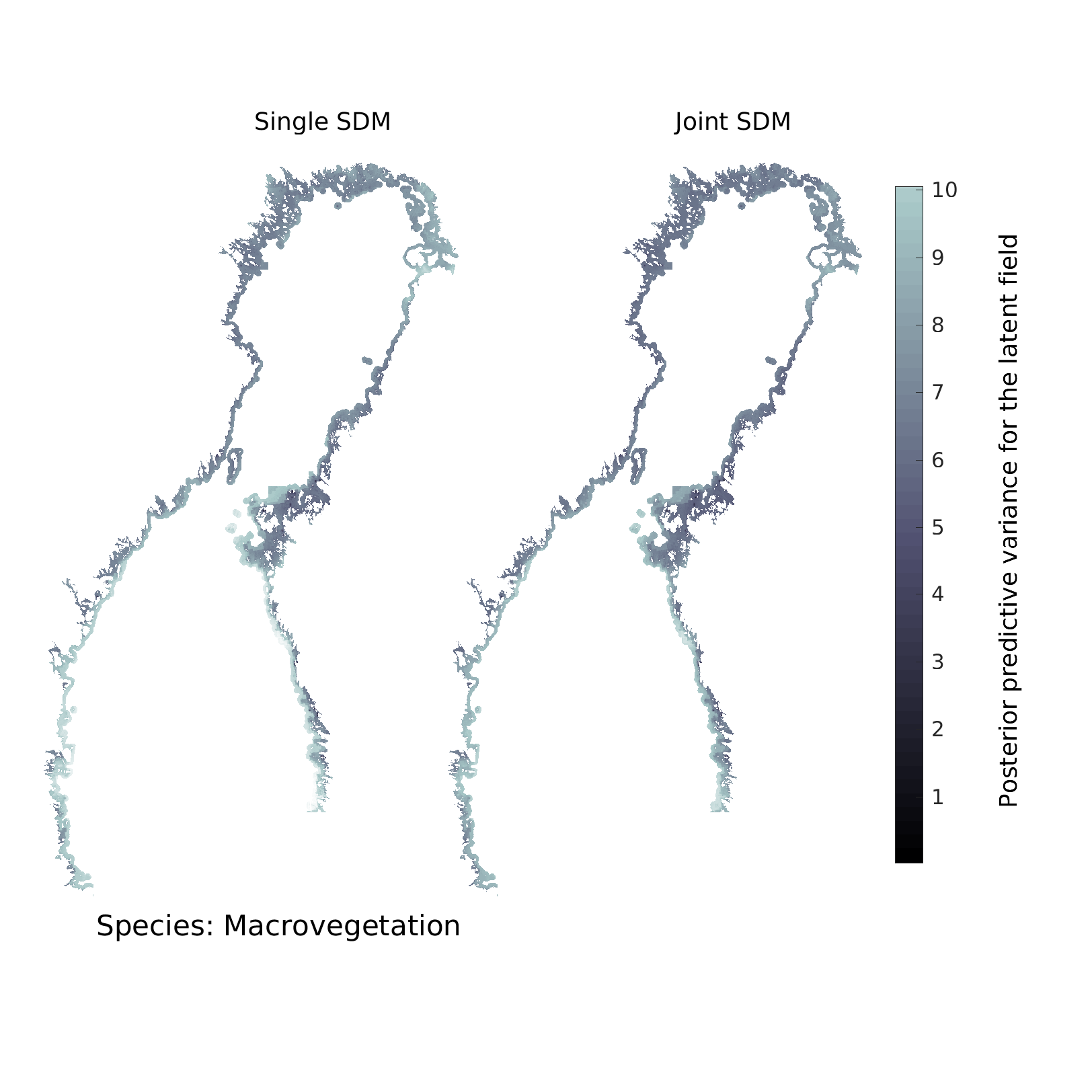}
\caption{Posterior predictive variance for the latent field $f_*(\x, \s)|\y, \hat{\eta}, \hat{\lambda}$ for macrovegetation predicted by SSDM (model 1) and JSDM (model 3).}
\end{figure}

\begin{figure}[!htb]
\includegraphics[scale=0.75]{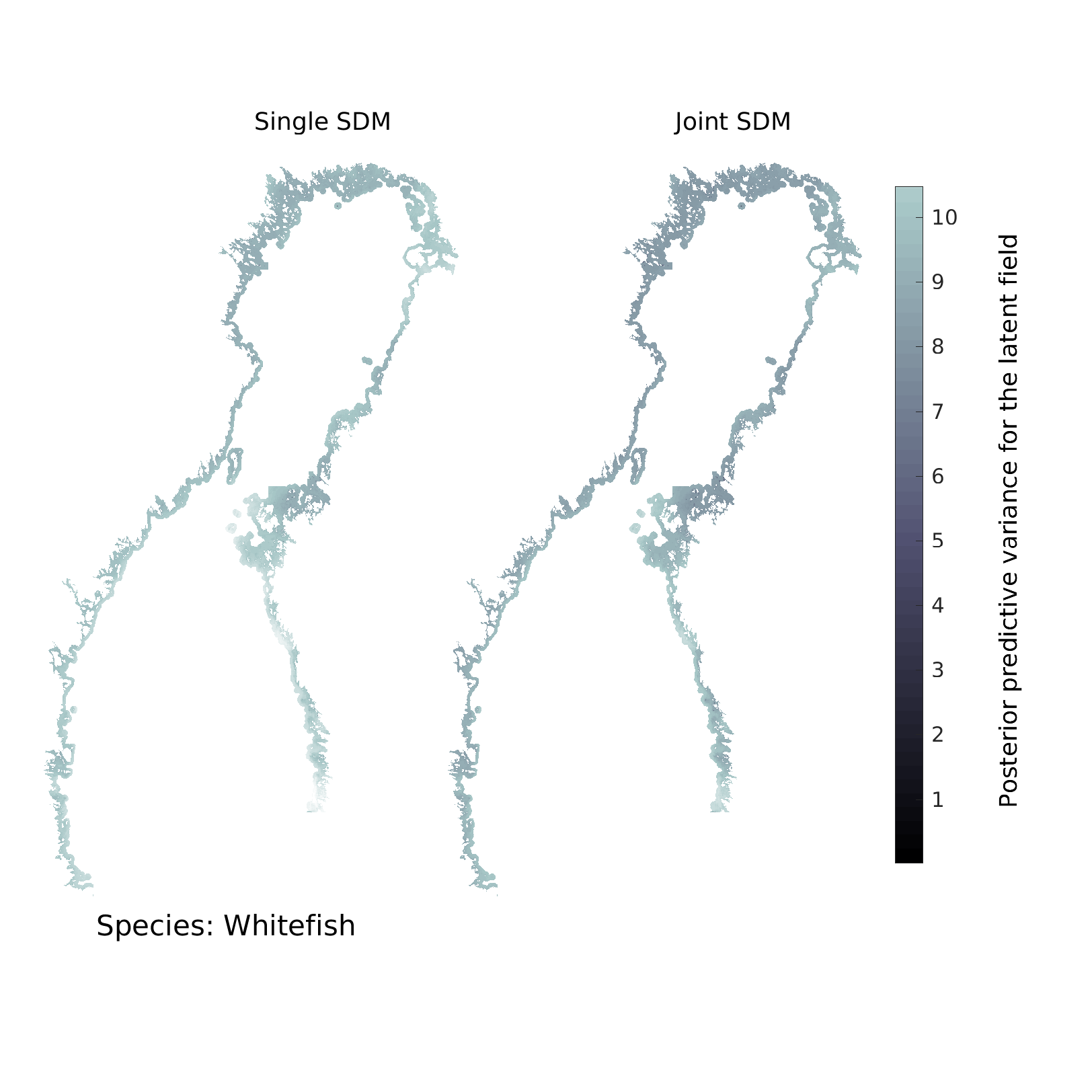}
\caption{Posterior predictive variance for the latent field $f_*(\x, \s)|\y, \hat{\eta}, \hat{\lambda}$ for whitefish predicted by SSDM (model 1) and JSDM (model 3).}
\end{figure}

\begin{figure}[!htb]
\includegraphics[scale=0.75]{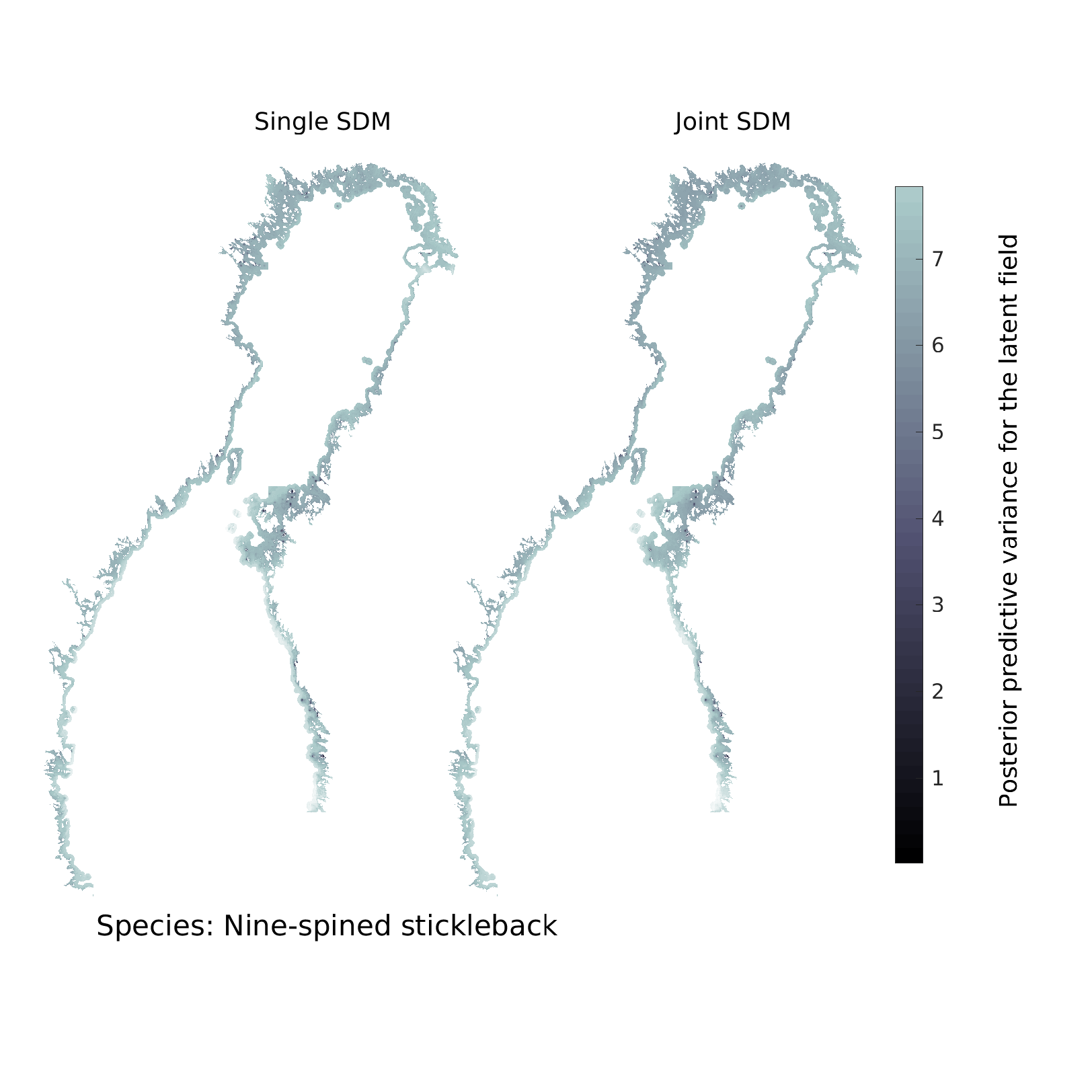}
\caption{Posterior predictive variance for the latent field $f_*(\x, \s)|\y, \hat{\eta}, \hat{\lambda}$ ninespine-stickleback predicted by SSDM (model 1) and JSDM (model 3).}
\end{figure}
\begin{figure}[!htb]
\includegraphics[scale=0.75]{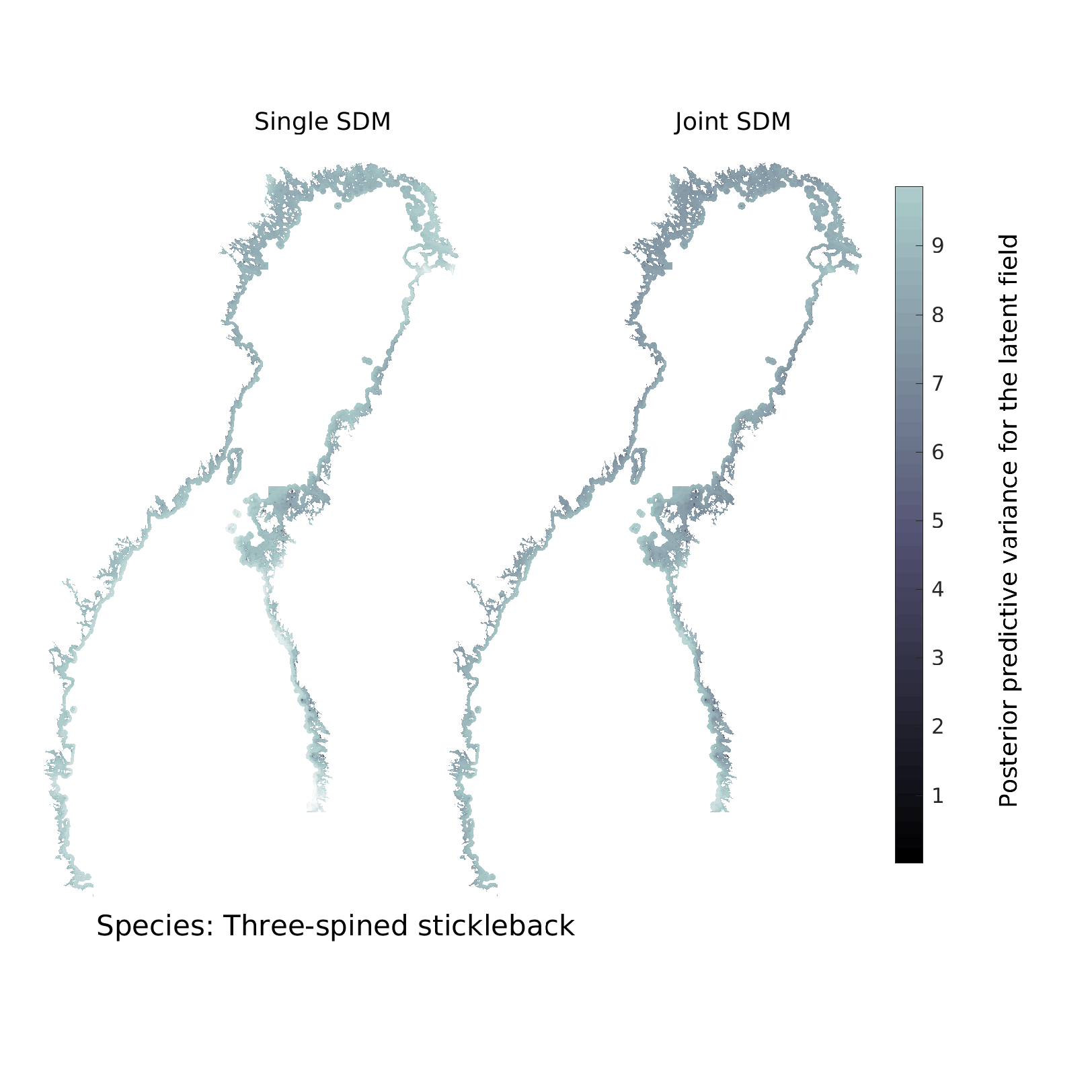}
\caption{Posterior predictive variance for the latent field $f_*(\x, \s)|\y, \hat{\eta}, \hat{\lambda}$ for threespine-stickleback predicted by SSDM (model 1) and JSDM (model 3).}
\end{figure}